\documentclass[preprint,12pt, authoryear]{elsarticle}
\usepackage[utf8x]{inputenc}
\usepackage{blindtext} 
\usepackage{color}
\usepackage{mathtools} 
\usepackage{lmodern,amsmath,amssymb}
\usepackage[titletoc,title]{appendix}
\usepackage{bm}
\usepackage{nomencl}
\usepackage{multirow}
\usepackage{siunitx} 
\usepackage{subcaption} 
\usepackage{geometry}
\usepackage{rotating}
\usepackage{setspace}

\renewcommand{\vec}[1]{{\boldsymbol{\mathbf{#1}}}}
\usepackage[font={normalsize}]{caption}
\usepackage{etoolbox,ragged2e}
\newcommand{\p}{\mathrm{p}}
\newcommand{\f}{\mathrm{f}}
\newcommand{\col}{\mathrm{c}}
\newcommand{\g}{\mathrm{g}}
\newcommand{\lub}{\mathrm{lub}}
\newcommand{\tot}{\mathrm{t}}

\begin{document}
\title{ A collision model for DNS with ellipsoidal particles in viscous fluid}
\author[tudd]{Ramandeep Jain \corref{cor1}}
\author[ilk]{Silvio Tschisgale}
\author[tudd]{Jochen Fröhlich}
\address[tudd]{Institute of Fluid Mechanics, Technische Universität Dresden \\
	George-Bähr Str. 3c, D-01062 Dresden, Germany \\}
\address[ilk]{Institut für Luft- und Kältetechnik Dresden, 01309 Dresden, Germany}
\cortext[cor1]{Corresponding author: ramandeep.jain@tu-dresden.de}
\doublespacing
\begin{abstract}
	The article proposes an algorithm to model the collision between arbitrary ellipsoids in viscous fluid. It is composed of several steps, each improving upon the standard procedure employed in the current literature. First, an efficient contact detection algorithm is presented. Then, the semi-implicit Immersed Boundary Method of Tschisgale et al., $2018$a (A general implicit direct forcing immersed boundary method for rigid particles, {\em Computers \& Fluids}, $170$:$285–298$) is enhanced by the collision forces using the hard-sphere approach, so that the resulting model accounts for fluid forces throughout the entire collision process. Additionally, a new lubrication model is proposed that applies a constant lubrication force in the region where hydrodynamic forces are not resolved by the spatial grid. The new collision model is validated against benchmark test cases of particle wall collisions with excellent agreement. Furthermore, the collision of an ellipsoidal particle with a wall is investigated. The normal restitution coefficient in the case of ellipsoidal particles does not solely depend upon the Stokes number as in the case of spherical particles but is also a function of its shape, and the orientation before the collision. Using the new model, the effect of these parameters on the rebound trajectory is studied. It is found that the maximum normal restitution coefficient decreases significantly as the flatness of the particle increases. Also, the coefficient of restitution depends on the particle orientation, a tendency increasing with particle flatness.
\end{abstract}
\maketitle

\newpage
\section{Introduction}
Flows laden with particles are of common occurrence in many applications in process engineering, geophysics, river engineering, and oceanography. Fluidized bed reactors, the sediment transport in rivers, sedimentation in sewage treatment plants, and turbidity currents are only some prime examples. The fluid fields of such flows are substantially modified by the presence of the particles if their size is larger than the Kolmogorov length scale \textcolor{red}{\citep{brennen2005,voth2017}}. This influence is further enhanced with a higher particle volume fraction as interparticle interactions become important in such cases \citep{Elghobashi1994,brennen2005}. A classical example is the flow over and through a dense, mobile granular bed, as in rivers for example.

In the recent past, investigations of the fluid-particle and particle-particle interaction in such a system have drastically changed due to the availability of accurate and numerically efficient simulations and due to the ever increasing computing power. Many numerical methods to conduct such simulations have been developed in the last few years, as described in \cite{Uhlmann2005, derksen2011, kempe2012b, Kidanemariam2014a, Sun2016, Ardekani2016, biegert2017,  Rettinger2017b}, and others. A direct numerical simulation (DNS), where the smallest turbulent length scale is resolved, coupled with immersed boundary method (IBM) is one of the most frequently implemented and promising frameworks \citep{Mittal2005}. Employing such an algorithm for densely laden flows, the interparticle collision is generally resolved using a soft-sphere collision model \citep{Crowe1998,kempe2012a,Kidanemariam2014a,Izard_2014b,Ardekani2016,biegert2017}. These tools have provided many important insights in to the physics of particle-laden open channel flows  \citep{kempe2014,Derksen2015, vowinckel2016, kidanemariam_2017,Jain2017,biegert2017}.

Most of the previously conducted numerical simulations investigating the transport of heavy particles in turbulent open-channel flow considered spherical particles to represent the disperse phase. Natural sediment, however, exhibits non-spherical shapes, and various experiments pointed out major differences in particle movement when changing the particle shape \citep{Krumbein1941,Lane1954,Bradley1972,Rice1991}. Moreover, many authors argued that the collisions between non-spherical particles can be elastic or inelastic depending upon the shape and orientation of the colliding particle \citep{Abbott1977a,drake1988,Nino1994,schmeeckle2001}. In most of these experiments, saltation of an individual particle was photographed but the detailed observation of particle-bed interaction has led to different conclusions. \citet{Abbott1977a} stated that the collisions are viscously damped and inelastic, whereas \citet{Nino1994} observed that some of the particles rebound with part of the collisions being elastic. \cite{schmeeckle2001} found that many particles did not rebound even if the Stokes number $St_\mathrm{r}$ was significantly higher than the threshold value of $10$, above which a spherical particle rebounds according to the experiments of \citet{gondret2002}. 

\citet{schmeeckle2001} also observed the orientation of natural sediment particles colliding with an inclined glass wall in water. They noticed that the orientation of the particle before collision was an important parameter to determine the normal restitution coefficient. In fact, it is crucial to take into account whether the collision force is directed through the center of mass of the particle, or not. Despite a rebound of the contact point, the center of mass of the particle may not rebound because of the non-spherical shape of the natural sediments. The majority of the natural particles did not rebound in the cited experiment because of the off-center collisions. Such inelastic or viscously damped collisions provide less impact to set another particle in motion, which is an important parameter in an erosion event \citep{vowinckel2016}. The experience reported in the literature, hence, boulders the importance of addressing the particle shape when studying sediment transport in turbulent channel flow.

Representing an arbitrary shape does not pose any difficulty for the DNS-IBM framework to account for the particle-fluid interaction, and \citet{tschisgale2018} recently proposed an improved IBM for this purpose. The difficulty lies in modeling the inter-particle collisions and the contribution of the surrounding fluid. In case of partly concave shapes, as observed with some types of natural grains, multiple contact points are possible. One way to account for complex shapes is to represent these by an arrangement of spheres, thus reverting to the easier shape and using the related algorithms. This was done by \citet{Fukuoka2014} and \citet{Sun2017}, for example, and is a common strategy in DEM methods (e.g. \cite{Favier1999}, \cite{Mede2018}). 

Another approach is to rule out non-convex shapes and to represent the particles with convex shapes. Different kinds of representations have been used in such simulations, like cube \citep{Rahmani2014}, cylinder \citep{Dorai2015}, tetrahedron \citep{Rahmani2014} etc. When treating sediment in the literature on oceanography and geology it is common practice to characterize particles according to the definition of \citet{Zingg1935} by measuring their major, intermediate, and minor axis. The Zingg representation is considered very suitable in the literature to represent the natural shape \citep{Allen1985}. Furthermore, this shape is strictly convex, numerically well treatable and, hence, selected here.

To conduct simulations with huge number of ellipsoidal particles, where billions of collisions are happening at a time, a robust and efficient algorithm for ellipsoids is required. \citet{Ardekani2016} recently presented a collision model for spheroidal particles in viscous media extending the linear spring-dashpot system developed for spherical particles. With this model, the short-range interactions due to lubrication forces are approximated by replacing the spheroids with spheres of equal mass and a radius corresponding to the local radius of curvature at the contact point. A simplified model for non-spherical particles, where collision forces are applied only through a spring, was presented in \citet{Mohaghegh2017}. \textcolor{red}{\cite{Saini2019} recently used a non-linear spring-dashpot-slider system without any lubrication model to simulate the collision between two ellipsoids in shear flow.} While there is not much literature related to the collision model for ellipsoids in viscous fluid, the granular flow and the computer graphics communities have been working on collisions between ellipsoids in dry environment since more than a decade \citep{Guendelman2003,Tonge_2012,Buist2016}. Using these algorithms without alteration, however, is problematic since collisions are notably different in viscous media.

Most of the collision models developed for spherical particles in fluid are based on a linear soft-sphere system as, e.g., used by \citet{Izard_2014b,Kidanemariam2014a} and \citet{Ardekani2016}. Other implementations are based on the Herzian contact theory, which is nonlinear, as in \citet{kempe2012a}, \citet{Ray2015}, and \citet{biegert2017}. A drawback of a soft-sphere model is the reduced realism if, for reasons of stability, the stiffness of the collision is reduced, often by orders of magnitude, such that an uncontrolled interpenetration occurs. This can be reduced to a minimum by an optimization process that leads to the Adaptive Collision Model (ACM) proposed by \citet{kempe2012a} with applications in \citet{kempe2014,vowinckel2016} and others. Another approach is sub-stepping in time for the particles alone as performed by \citet{Kidanemariam2014a} and \citet{biegert2017}.

In contrast to soft-sphere models, so called Hard-sphere models represent the collision with infinitely short collision time. This is realistic since characteristic time scales during direct particle contact are shorter by about two orders of magnitude compared to the time scales of particle-fluid interaction for realistic material properties \citep{kempe2012a}. Furthermore, hard-sphere models do not contain any numerical parameter which may deteriorate physical realism and needs to be adjusted. Multiple collision, however, are dramatically more complex and require special treatment. In the computer graphics community, dynamics involving multiple contact are formulated as a linear complementarity problem (LCP), and the collision response is determined in an iterative way \citep{Guendelman2003,Tonge_2012} using, e.g., the projected Gauss-Seidel method. In this way, the multiple contact problem is treated as a sequence of single collisions between pairs of two particles, where iterations over these pairs are performed until all constraints are fulfilled.

As in some cases the somewhat lower degree of physical realism of the collision model can be accepted or compensated, e.g. by sub-stepping, particle-resolving simulations of sediment transport are most often conducted with soft-sphere models. So far, only very few studies employed a hard-sphere model for the simulations of particles in viscous flow \citep{derksen2011,Derksen2015,Rettinger2017}. \citet{Buist2016} argued that the efficiency of the hard-sphere model decreases with increasing number of simultaneous collisions. Therefore, they proposed a hybrid collision model composed of a hard-sphere model for solving binary collisions and a soft-sphere model for multibody collisions. Recently, \citet{Rettinger2017} employed a hard-sphere collision to conduct simulations of dune formation. To reduce the effort, they avoided multiple contacts by imposing that only collision pairs are treated and that the number of colliding pairs equals the number of particles. This choice is made by a contact detection algorithm executed before treating the collisions. 

The available collision models in one way or another all suffer from sacrificed physical realism or from increased numerical complexity, both introduced to limit the computational effort. Furthermore, the collision models so far applied to sediment transport and beyond, treat the collision process as a "dry collision" accounting for fluid forces only before and after the collision, but not during the collision itself. This has dramatic consequences in case of a hard-sphere model in that fluid forces can become very large for realistic time stepping as experienced in \citet{kempe2012a} where the observation triggered the development of the ACM, which is an optimized self-adaptive soft-sphere model. 

The present paper proposes a hard-sphere model exempt of the above restrictions with fluid forces incorporated during the collision. Here, the term collision is understood to cover the entire phase of particle-particle interaction, composed of the phase of direct surface contact as well as the phase of very small surface distances not resolved by the computational grid, where lubrication forces dominate. For sediment transport the collision model also has to cover the situation of sustained contact with particles constantly exerting forces, due to gravity, for example. When employing a soft-sphere model this requires particular measures, as described in \cite{kempe2012a,Kidanemariam2014a,biegert2017}. With the present approach sustained contact is covered realistically without any special treatment. The development presented here builds upon the collision model for Cosserat rods recently proposed by \citet{Tschisgale2018b} and the semi-implicit IBM of \cite{tschisgale2018}. 

The originality of the article comes from these new features. 1) A very efficient contact detection algorithm, 2) introducing the semi-implicit IBM in the collision model, 3) a new lubrication model which is versatile, robust, and efficient. The resulting model is validated against experimental data achieving very satisfactory agreement. Later, the model is used to explore the effect of particle shape and orientation on its rebound trajectory after a collision.

\section{Governing equations}
\subsection{Continuous phase}
\begin{figure}
	\hspace*{0.2\linewidth}
	\def\svgwidth{0.6\linewidth}
	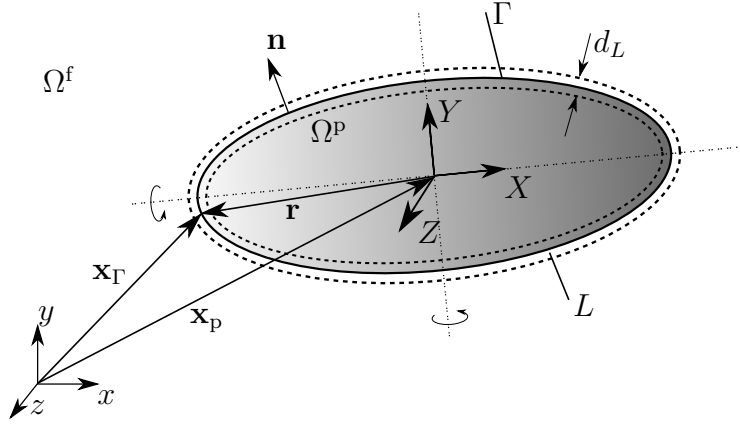
	\caption{An instantanous snapshot of an ellipsoid as a moving particle. The nomenclature is introduced in the text.}
	\label{fig:sketch_domain}
\end{figure}

The fluid field is described by the unsteady three dimensional Navier-Stokes equations for a Newtonian fluid of constant density, 
\begin{equation} 
 \frac{\partial{\vec{u}}}{\partial{t}} = - \nabla\cdot(\vec{u}\otimes\vec{u}) + \frac{1}{\rho_\f} \: \nabla\cdot\boldsymbol{\tau}+\vec{f}_\mathrm{v}+\vec{f}_\mathrm{IBM} \hspace{0.1cm},
\end{equation} 
%
\begin{equation}
\nabla\cdot\vec{u}=0     \hspace{0.1cm} .
\end{equation}
which are solved on an equidistant Eulerian grid in the domain $\Omega = \Omega^{\f} \cup \Omega^{\p}$, i.e. the union of the fluid domain and the particle interior (Fig. \ref{fig:sketch_domain}). This nomenclature differs from \cite{tschisgale2018}. Here, $\textbf{u}=(u,v,w)^{T}$, $p$, $\rho_\f$, and $t$ are the velocity vector along the Cartesian coordinates $x,y,z$, pressure, fluid density, and time, respectively. On the right-hand side, $\vec{f}_\mathrm{IBM}$ is the coupling force imposed to take into account the particle-fluid interaction, and $\vec{f}_\mathrm{v}$ represents the volume force applied to drive the flow. The latter is constant in space and adjusted in time to obtain the desired flow rate. The hydrodynamic stress tensor, $\boldsymbol{\tau}$, is 
\begin{equation}
 \boldsymbol{\tau}= -p \:\vec{\mathbb{I}} + \mu_\f \:(\nabla \:\vec{u} + \left(\:\nabla \:\vec{u} \:\right)^\top ) \hspace{0.1cm},
\end{equation}
with $\mu_\f$ the dynamic viscosity of the fluid, and $\vec{\mathbb{I}}$ the identity matrix.\\

\subsection{Disperse phase}
The solid phase is represented by a non-deformable ellipsoid defined by its longest axis, $a$, intermediate axis, $b$, and smallest axis, $c$, as shown in Fig. \ref{fig:sketch_domain}. Here, $\vec{x}_\mathrm{p}$ is the position of particle center, $\vec{x}_\mathrm{\Gamma}$ an arbitrary point on the particle surface $\Gamma$, and $\vec{r} = \vec{x}_\mathrm{\Gamma}-\vec{x}_\mathrm{p}$ the vector connecting these points. The translational momentum balance of a particle is given by
\begin{equation}\label{eq:part_lin}
m_\p \:\frac{\text{d}\vec{u}_\p}{\text{d} t} =  \int_{\Gamma} \!\boldsymbol{\tau} \cdot \vec{n} \:\text{d} S   + V_\p\:( \rho_\p-\rho_\f ) \:(\vec{g}+\vec{f}_\mathrm{v}) + \vec{f}_\col + \vec{f}_\lub  \hspace{0.1cm},
\end{equation}
Here, $V_\p$, $\rho_\p$, and $m_\p$ are the volume, mass density, and mass of the ellipsoid, respectively, while $\textbf{u}_\p$ is the translational velocity of the particle, $\textbf{n}$ the outward-pointing surface normal vector, and $\textbf{g}$ the gravitational acceleration. The collision force $\vec{f}_\col$ and lubrication forces $\vec{f}_\lub$ are provided in the later sections below. The hydrodynamic loads are integrated over the particle surface $\Gamma$, which envelopes the particle domain $\Omega^{\p}$. The temporal evolution of the center of mass of the particle $\textbf{x}_\p$ is obtained by integrating
\begin{equation}\label{eq:part_lin_xp}
\frac{\text{d}\vec{x}_\p}{\text{d} t} = \vec{u}_\p \hspace{0.1cm}.
\end{equation}
The angular momentum balance is formulated in a body-fixed and co-rotated frame ${\mathcal{P}}$. This local coordinate system is defined to coincide with the principal axes of the rigid particle (Fig.~\ref{fig:sketch_domain}), such that the tensor of inertia $\textbf{I}_\p$ is diagonal, i.e.
\begin{equation}\label{eq:inertia_tensor}
\textbf{I}_\p=\text{diag}\left(\text{I}_1,\text{I}_2,\text{I}_3 \right)  \hspace{0.1cm}.
\end{equation}
The relation between the global and the local co-rotated frame is established by a rotation in terms of a unit quaternion $\textbf{q}_\p \in \mathbb{S}^3$, where $\mathbb{S}^3$ is the set of unit quaternions \citep{Kuipers1999}. It represents the spatial rotation of the particle around its center of mass $\textbf{x}_\p$. The temporal evolution of $\textbf{q}_\p$ is connected to the local angular velocity of the particle $\boldsymbol{\Omega}_\p$ via
\begin{equation}\label{eq:omega_quat}
\frac{\text{d} \textbf{q}_\p}{\text{d} t} = \frac{1}{2} \: \textbf{q}_\p \circ \boldsymbol{\Omega}_\p  \hspace{0.1cm},
\end{equation}
where $\dots$$ \circ \dots$ is the multiplication of two quaternions. Any vector $\vec{v}$ can be rotated from one coordinate system to the other using 
\begin{subequations}
	\begin{align}
	\label{eqn:forward_rot}
	\textbf{v} &= \textbf{q} \circ \textbf{V} \circ \bar{\textbf{q}} \hspace*{1cm} \text{(forward rotation)} \\
	\label{eqn:backward_rot}
	\textbf{V} &= \bar{\textbf{q}} \circ \textbf{v} \circ \textbf{q} \hspace*{1.09cm} \text{(backward rotation)}
	\end{align}
\end{subequations}
with $\bar{\textbf{q}}$ indicating the conjugate of a quaternion ${\textbf{q}}$.
As a convention, capital letters, e.g. $\textbf{V} \in \mathbb{R}^3$, are given in the local co-rotated frame and small letters, $\textbf{v} \in \mathbb{R}^3$, in the global Eulerian frame. Consequently, the angular velocity in the global Eulerian frame can be expressed in terms of $\boldsymbol{\Omega}_\p$, by
\begin{equation}\label{eq:transformation}
\boldsymbol{\omega}_\p = \textbf{q}_\p \circ \boldsymbol{\Omega}_\p \circ \bar{\textbf{q}}_\p  \hspace{0.1cm}.
\end{equation}
The angular momentum balance in the local coordinate system, then, reads
\begin{equation}\label{eq:part_ang}
\textbf{I}_\p \! \cdot \! \frac{\text{d} \boldsymbol{\Omega}_\p}{\text{d} t} + \boldsymbol{\Omega}_\p \times \textbf{I}_\p \!\! \cdot \! \boldsymbol{\Omega}_\p =  \int_{\Gamma} \textbf{R} \times \left( \textbf{T} \cdot \textbf{N} \right) \:\text{d} S + \vec{M}_\col + \vec{M}_\lub \hspace{0.1cm},
\end{equation}
where $\textbf{N}$ the outward pointing surface normal vector in the frame ${\mathcal{P}}$, $\textbf{R}$ the vector of a surface point with respect to the center of mass of the particle, $\textbf{T}$ the hydrodynamic stress tensor, $\vec{M}_\col$ the torque due to the collision and $\vec{M}_\lub$ the torque due to the lubrication force.
\subsection{Coupling between the two phases}
The coupling of fluid and particle equations is thoroughly discussed in \citet{tschisgale2018}. The final discretized model is recalled here and is the starting point for the derivation of the collision model. The discretized linear equation of motion of an ellipsoid coupled with a fluid using the IBM presented in that reference is
\begin{equation}\label{eq:lin_par}
\textbf{u}_\p^n - \textbf{u}_\p^{n-1} = \:\left( m_\p+m_L \right)^{-1} \: \Delta{t}\left\{ \vec{f}_\f + \vec{f}_\mathrm{g} + \vec{f}_\col + \vec{f}_\lub +\vec{f}_\mathrm{v} \right\} \hspace{0.1cm},
\end{equation}
where $m_{L}$ is the mass of the fluid layer ${L}$ of thickness $d_{L}$ constituting the central element in this method with all details and justifications provided in the cited reference. The superscript $n$ denotes the time step, while 
\begin{subequations}\label{eq:f_fandf_g}
\begin{align}
\vec{f}_\f &=- \: \rho_\f \! \int_L \:\frac{\textbf{u}_\Gamma^{n-1} - \tilde{\textbf{u}}^{} }{\Delta{t}}\: \text{d} V + \frac{1}{\Delta{t}} \left[\int_{\Omega^{\p}} \rho_\f \:\textbf{u} \:\text{d} V  \right]^{\:n}_{\:n-1}  \hspace{0.1cm},\\
\vec{f}_\mathrm{g} &= \: V_\p\:( \rho_\p-\rho_\f ) \:\textbf{g} \hspace{0.1cm},
\end{align}
\end{subequations}
are the time-discrete fluid and gravitational force, respectively. 
Here, $\textbf{u}_\Gamma$ is the velocity vector at a point on the surface of the moving particle and $\tilde{\textbf{u}}$ is the fluid velocity without taking into account the disperse phase.

Analogously, the angular equation of motion can be formulated as
\begin{equation}\label{eq:ang_par}
\boldsymbol{\Omega}_\p^n - \boldsymbol{\Omega}_\p^{n-1} = - \left( \textbf{I}_\p+\textbf{I}_L \right)^{-1} \left\{  \int^{n}_{n-1} \!\! \boldsymbol{\Omega}_\p \times \textbf{I}_\p \!\! \cdot \! \boldsymbol{\Omega}_\p \: \text{d} t \right\} + \Delta{t} \left( \textbf{I}_\p+\textbf{I}_L \right)^{-1} \! \cdot \! \left\{  \vec{M}_\f  + \vec{M}_\col + \vec{M}_\lub \right\} \hspace{0.1cm},
\end{equation}
with the torque due to the fluid forces, 
\begin{equation}\label{eq:T_f}
\begin{split}
\vec{M}_\f = & - \: \rho_\f \! \int_L \textbf{R}\times \left(\frac{\textbf{U}_\Gamma^{n-1} - \tilde{\textbf{U}}^{} }{\Delta{t}} \right) \text{d} V + \frac{1}{\Delta{t}} \left[\bar{\textbf{q}}_\p \! \circ \int_{\Omega^{\p}} \rho_\f \:\textbf{r}\times \textbf{u} \:\text{d} V \:\: \circ \textbf{q}_\p  \right]^{\:n}_{\:n-1} \hspace{0.1cm}.
\end{split}
\end{equation}
Here, $\textbf{I}_L$ is the tensor of inertia introduced by the layer $L$ shown in Fig. \ref{fig:sketch_domain}.
Considering Eq. \eqref{eqn:backward_rot}, $\textbf{U}_\Gamma$ and $ \tilde{\textbf{U}}^{}$ are the same as $\textbf{u}_\Gamma$ and $\tilde{\textbf{u}}$, respectively, but in the co-rotated frame $\mathcal{P}$ of the respective particle. The vector joining a point on the surface of an ellipsoid to the center of that ellipsoid in the local frame of reference is denoted as $\vec{R}$ with $\vec{R} = \bar{\textbf{q}} \circ \vec{r} \circ \textbf{q}$ and $\vec{r}$ the same vector in the global frame.
\section{Contact detection}
Any collision model has to consist of two steps, detection of collisions and determination of the interaction between the collision partners. Both can be implemented in various ways depending on the particular type of model. The detection of a collision generally involves finding the closest distance between two collision partners. For spherical particles this is elementary, as it can be deduced from the distance of the particle centers. For all other particle shapes the closest distance between points on the two surfaces depends on their orientation. If many particles are involved, as in the case of sediment transport, this step can become extremely expensive.

Recently, \citet{Ardekani2016} introduced an algorithm for spheroids, where two spheres are moved iteratively inside the particles until the slope of the line that connects the closest points is below a given threshold. It is mentioned that the convergence rate of this method depends on the size of the chosen sphere \citep{Ardekani2016}. In the present work, an iterative algorithm is provided that is exempt of such an issue, converges fast and is free from any additional parameter. 

\begin{figure}
	\centering
	\def\svgwidth{0.6\linewidth}
	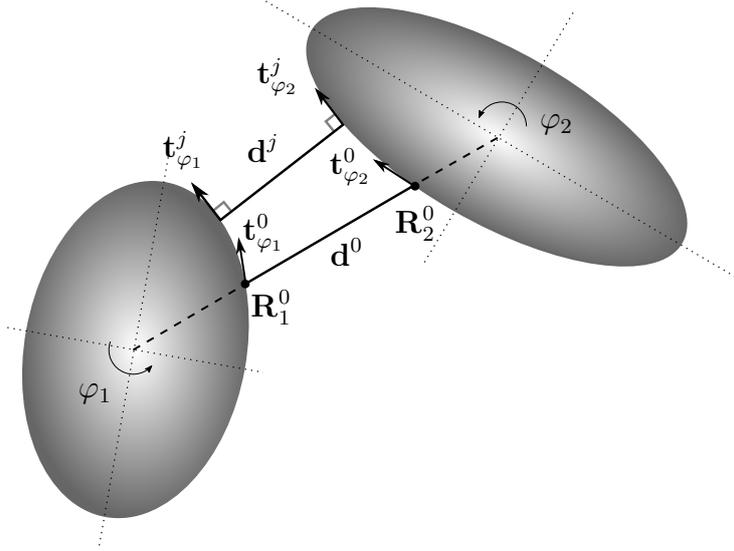
	\caption{Two approaching ellipsoids and the variables used in the algorithm to find the closest distance between these two ellipsoids.}
	\label{fig:contact_detection}	
\end{figure}

A point on the surface of an ellipsoid is represented as $\vec{R} = (R_X, R_Y, R_Z)^T$ with $R_X$, $R_Y$, and $R_Z$ the point coordinates in the body-fixed coordinate system $\mathcal{P}$, as displayed in Fig. \ref{fig:contact_detection}. This algorithm uses the parametric equations of an ellipsoid
\begin{equation} 
\begin{split}
	R_X \quad & = \quad a \, \cos(\varphi) \, \sin(\vartheta) \\
	R_Y \quad & = \quad b \, \sin(\varphi) \, \sin(\vartheta) \\
	R_Z \quad & = \quad c \, \cos(\vartheta)
\end{split}
\end{equation}

with $\vartheta \in [0,\pi]$ and $\varphi \in (-\pi,\pi]$ the polar and azimuthal angle, respectively. By differentiating these equations in $\vartheta$- and $\varphi$-direction, the tangent vectors $\vec{t}_{\vartheta_i}$ and $\vec{t}_{\varphi_i}$ are found at a point $\vec{R}_{i} = \vec{R}_{i}(\vec{a_i}, \vartheta_i,\varphi_i)$ on the surface of the ellipsoid with index $i$ having the axes $\vec{a}_i=(a_i,b_i,c_i)^T$. Here, $a_i$, $b_i$, and $c_i$ are the longest, intermediate, and the smallest axis of the ellipsoid, respectively. The vector $\vec{d}$ is the distance vector joining the point $\vec{R}_{1} = \vec{R}_{1}(\vec{a}_1, \vartheta_1, \varphi_1)$ on the surface of ellipsoid $1$ with $\vec{R}_{2}= \vec{R}_{2}(\vec{a}_2, \vartheta_2$, $\varphi_2)$ on ellipsoid $2$. With $\vec{a}_1$ and $\vec{a}_2$ fixed the two points are a function of the angles $(\vartheta_1, \varphi_1)$ and $(\vartheta_2, \varphi_2)$, respectively. The method used in this algorithm exploits the fact that the shortest distance vector must be perpendicular to the tangent vectors. 

To this end an iterative process is established, updating $(\vartheta_1, \varphi_1)$ and $(\vartheta_2, \varphi_2)$ simultaneously using 
\begin{subequations}
\begin{align}
\vartheta_1^{j+1} &= \vartheta_1^j + \frac{D_\mathrm{eq,1}}{a_1} \, \frac{ \vec{d}^j \!\! \cdot \! \vec{t}_{\vartheta_1}^j }{\lvert \vec{d}^j \rvert \, \lvert \vec{t}^j_{\vartheta_1} \rvert} \hspace{0.1cm}, \\
\varphi_1^{j+1} &= \varphi_1^j + \frac{D_\mathrm{eq,1}}{a_1} \, \frac{ \vec{d}^j \!\! \cdot \! \vec{t}_{\varphi_1}^j }{\lvert \vec{d}^j \rvert \, \lvert \vec{t}^j_{\varphi_1} \rvert} \hspace{0.1cm},
\end{align} \label{eq:contact_detection}
\end{subequations}
for ellipsoid $1$ and the same expression with the index $1$ replaced by $2$ for ellipsoid $2$. Here, $D_\mathrm{eq,1}$ is the volumetrically equivalent diameter of ellipsoid $1$.
The superscript $j$ represents the $j$-th iteration. The initial surface points $\vec{R}^0_{1}$ and $\vec{R}^0_{2}$ are obtained by taking the surface points lying on the line connecting the centers of the two ellipsoids (Fig. \ref{fig:contact_detection}). The iterative procedure is terminated when the dot products $\vec{d}^j\! \cdot  \vec{t}^j_{\vartheta_1} $ and $ \vec{d}^j \! \cdot   \vec{t}^j_{\varphi_1} $ vanish or, technically, are smaller than a given threshold. With \eqref{eq:contact_detection} the change in $\vartheta_1$ and $\varphi_1$ is proportional to the dot product to increase the rate of convergence. The factor $D_\mathrm{eq}/a$ is introduced to account for regions of increased curvature encountered with stronger non-sphericity. The collision with a wall is modeled as a collision with a sphere of radius $R = 10^{12}$.  Two objects are considered to be in contact if the converged distance fulfills $\lvert \vec{d} \rvert < \epsilon_c$, with the margin for contact $\epsilon_c = 0.2\Delta_x$ and $\Delta_x$ the stepsize of the equidistant Eulerian fluid grid.

The present algorithm is started only if the surfaces of the two spheres enveloping the ellipsoids are sufficiently close to each other, i.e. $\lvert \vec{x}_\mathrm{p,1} - \vec{x}_\mathrm{p,2}\rvert \leq a_1+a_2+d_\mathrm{lub}$. The additional distance is chosen to be $d_\mathrm{lub} = 2\Delta_{x}$ and is related to the evolution of the lubrication model proposed in Sec. \ref{sec:lubrication_model} below. This avoids the unnecessary computation of the precise distance and reduces the computational effort in the overall contact detection procedure substantially.
\section{{\label{sec:collision_model}}Collision model}
\subsection{Single collision}
\label{sec:single_collision}
\begin{figure}
	\centering
	\def\svgwidth{0.7\linewidth}
	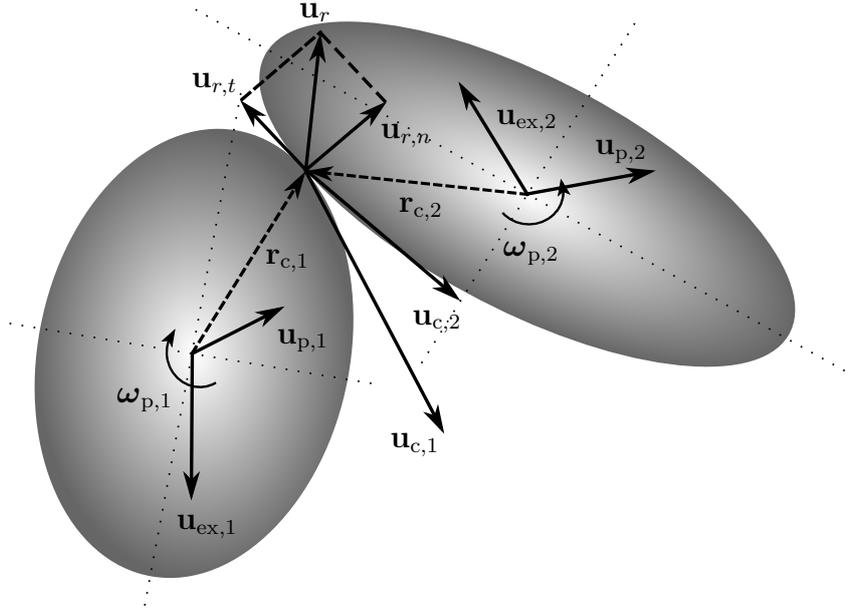
	\caption{Two ellipsoids in collision and the variables defining the system. The time index is dropped here for better readability. }
	\label{fig:colliding_pair}	
\end{figure}

Once it is found that the two ellipsoids are in contact, the collision response of both particles needs to be computed. The associated collision model used for this purpose is based on the model developed by \citet{Tschisgale2018b} for the collision of rod structures in dry environment. In the present work, this algorithm is enhanced to model the collision of rigid ellipsoids colliding in viscous fluid. This section describes the basic ideas of the collision modeling and the derivation of the collision forces and torques, $\vec{f}_{\mathrm{c}}$ and $\vec{m}_{\mathrm{c}}$, respectively. 
Staring point of the derivation is the configuration of two colliding ellipsoids at the moment of their first contact, shown in Fig. \ref{fig:colliding_pair}. 

To describe the time-discrete equations of motion of the particles, it is assumed that the instant shown in this figure coincides with the discrete time level $t^{n-1}$. Due to the high rigidity of the particles considered here, the timescale of contact is much smaller than the long-term behavior of the particles. Especially for perfectly rigid bodies, the collision time is infinitesimally small, so that at the moment of first touch, the velocities of the particles discontinuously change while the particle position and orientation remain unchanged. This implies that the collision force at the contact point has a Dirac-like impulsive nature. In the present time-discrete context, this collision force is approximated by a constant value acting over the time interval $\Delta t$ denoted $\vec{f}_{\mathrm{c}}$. The velocities of the particles after the collision process are then given at the new time level $t^{n}$.\\
The force $\vec{f}_{\mathrm{c}}$ can be derived taking advantage of the velocities at the contact point of both ellipsoids $\vec{u}_{\mathrm{c,1}}$ and $\vec{u}_{\mathrm{c,2}}$. After the collision, at time $t^{n}$, these are
\begin{equation}{\label{eq:velAtContact}}
\vec{u}^{n}_{\mathrm{c}} = \vec{u}^{n}_\p + \vec{\omega}^{n}_\p \times \vec{r}_{\mathrm{c}} \hspace{0.1cm},
\end{equation}
with the vector, $\vec{r}_{\mathrm{c}}$ joining the contact point $\vec{x}_\mathrm{c}$ and the particle center $\vec{x}_\mathrm{p}$. The index identifying the particle being dropped for ease of notation. In Eq. \eqref{eq:velAtContact}, the linear and angular velocity $\vec{u}^{n}_\p$ and $\vec{\omega}^{n}_\p$ can be replaced by utilizing the discretized equations of motion, \eqref{eq:lin_par} and \eqref{eq:ang_par}, so that
\begin{equation}{\label{eq:longlonglilong}}
\begin{split}
\vec{u}^{n}_{\mathrm{c}} & =  \textbf{u}_\p^{n-1} + \: m_\tot^{-1} \: \Delta{t} \: \left\{ \vec{f}_\f + \vec{f}_\mathrm{g} + \vec{f}_\col + \vec{f}_\lub \right\} \\
& + \left[ \textbf{q}_\p^{n-1} \circ \left( \boldsymbol{\Omega}_\p^{n-1} + \Delta{t} \: \textbf{I}_\tot ^{-1} \! \cdot \! \left\{  \vec{M}_\f  + \vec{M}_\col + \vec{M}_\lub \right\} - \textbf{I}_\tot ^{-1} \int^{n}_{n-1} \!\! \boldsymbol{\Omega}_\p \times \textbf{I}_\p \!\! \cdot \! \boldsymbol{\Omega}_\p \: \text{d} t \right) \circ \bar{\textbf{q}}_\p^{n-1} \right] \times \vec{r}_{\mathrm{c}} \hspace{0.1cm},
\end{split}
\end{equation}
with $m_\tot = m_\p + m_{L}$ and $\textbf{I}_\tot = \textbf{I}_\p + \textbf{I}_{L}$. After rearranging, this equation can be expressed in more compact form as
\begin{equation}{\label{eq:afterRearrange}}
\vec{u}^{n}_{\mathrm{c}} = \vec{u}^{n-1}_{\mathrm{c}} + \vec{u}_{\mathrm{ex}} + \: m_\tot^{-1} \: \vec{p}_\col 
+ \: \textbf{i}_\tot ^{-1} \! \cdot \! \vec{l}_\col \: \times \vec{r}_{\mathrm{c}} \hspace{0.1cm},
\end{equation}
where $\vec{u}^{n-1}_{\mathrm{c}}$ is the velocity at the contact point before the collision. Furthermore, $\vec{p}_\col = \Delta t \, \vec{f}_\col$ is the linear momentum due to the collision and $\vec{l}_\col = \Delta t \, \vec{m}_\col$ the angular momentum. The inertia matrix in the global coordinate system is obtained by the transformation $\textbf{i}_\tot = \mathcal{R} \cdot \textbf{I}_\tot \cdot \mathcal{R}^{\mathrm{T}}$, with the rotation matrix $\mathcal{R}$ converting any vector $\vec{s}$ from the body-fixed frame into the global coordinate system, i.e. $\mathcal{R} \cdot \vec{s} = \textbf{q}_\p \circ \vec{s} \circ \bar{\textbf{q}}_\p$. Finally, comparing Eqs. \eqref{eq:afterRearrange} and \eqref{eq:longlonglilong} shows that
\begin{equation}{\label{eq:externalForce}}
\begin{split}
\vec{u}_{\mathrm{ex}}  & = \Delta{t} \: m_\tot^{-1} \: \left\{ \vec{f}_\f + \vec{f}_\mathrm{g} + \vec{f}_\lub \right\} \\
                       & + \left[ \textbf{q}_\p^{n-1} \circ  \Delta{t} \: \textbf{I}_\tot ^{-1} \! \cdot \! \left\{  \vec{M}_\f  + \vec{M}_\lub \right\}  \circ \bar{\textbf{q}}_\p^{n-1} \right] \times \vec{r}_{\mathrm{c}} \\
                       & - \left[ \textbf{q}_\p^{n-1} \circ \left( \textbf{I}_\tot ^{-1} \int^{n}_{n-1} \!\! \boldsymbol{\Omega}_\p \times \textbf{I}_\p \!\! \cdot \! \boldsymbol{\Omega}_\p \: \text{d} t \right) \circ \bar{\textbf{q}}_\p^{n-1} \right] \times \vec{r}_{\mathrm{c}}
\end{split}
\end{equation}
which can be computed fully explicitly at $t^{n-1}$ before the collision since $\vec{f}_\f$ and $\vec{f}_\g$ are known from \eqref{eq:f_fandf_g}, provided that $\vec{f}_\lub$ and $\vec{M}_\lub$ are known. Such explicit expressions will be provided below. The quantity $\vec{u}_\mathrm{ex}$ in \eqref{eq:externalForce} represents the velocity induced by external forces including hydrodynamic, gravitational and lubrication forces acting during the collision time interval. It does not have a time index since $\vec{u}_{\mathrm{ex}}$ contains forces that have intermediate values and can not be associated to one time step. Note that the last term of Eq. \eqref{eq:externalForce} vanishes since
\begin{equation}
  \boldsymbol{\Omega}_\p \times \textbf{I}_\p \!\! \cdot \! \boldsymbol{\Omega}_\p = \bar{\textbf{q}}_\p \circ \frac{\text{d}\, \textbf{i}_\p}{\text{d}t} \vec{\omega}_\p \circ \textbf{q}_\p
\end{equation}
and ${\text{d}\, \textbf{i}_\p}/{\text{d}t} \to 0$ due to the assumption of infinitesimally short collision time \citep{Guendelman2003, Tschisgale2018b}.\\ 
In Eq. \eqref{eq:afterRearrange}, the angular momentum in the global Eulerian frame, $\vec{l}_\col$, is related to the linear momentum $\vec{p}_\col$ by 
$\vec{l}_\col = \vec{r}_{\mathrm{c}} \times \vec{p}_\col $ since there is only one contact point for colliding convex rigid bodies such as ellipsoids. This property allows to express Eq. \eqref{eq:afterRearrange} in the form
\begin{equation}{\label{eq:relVelFinal}}
\vec{u}^{n}_{\mathrm{c}} = \vec{u}^{n-1}_{\mathrm{c}} + \vec{u}_{\mathrm{ex}} + \: \vec{K} \cdot \vec{p}_\col \hspace{0.1cm},
\end{equation}
with the symmetric system matrix
\begin{equation}
\vec{K} =  m_\tot^{-1}\mathbb{I} + [\vec{r}_{\mathrm{c}}]^\mathrm{T}_\mathrm{x} \cdot \textbf{i}_\tot ^{-1} \cdot [\vec{r}_{\mathrm{c}}]_\mathrm{x} \hspace{0.1cm}.
\end{equation}
The term $[\vec{r}_{\mathrm{c}}]_\mathrm{x}$ is the skew matrix of the vector $\vec{r}_{\mathrm{c}}$ and is defined by the equality
\begin{equation}
[\vec{r}_{\mathrm{c}}]_\mathrm{x} \cdot \textbf{p}_\col = \vec{r}_{\mathrm{c}} \times \textbf{p}_\col \hspace{0.1cm}.
\end{equation}

The quantities in Eq. \eqref{eq:relVelFinal} differ for both colliding particles, except the shared collision load acting at the common contact point, so that 
\begin{subequations}
	\begin{align}
	\label{eq:relVelFinal1}
\vec{u}^{n}_{\mathrm{c,1}} &= \vec{u}^{n-1}_{\mathrm{c,1}} + \vec{u}_{\mathrm{ex,1}} - \: \vec{K}_1 \cdot \vec{p}_\col \hspace{0.1cm} \\
  \label{eq:relVelFinal2}
\vec{u}^{n}_{\mathrm{c,2}} &= \vec{u}^{n-1}_{\mathrm{c,2}} + \vec{u}_{\mathrm{ex,2}} + \: \vec{K}_2 \cdot \vec{p}_\col \hspace{0.1cm}
	\end{align}
\end{subequations}
Here, the linear momentum transferred in a collision, $\vec{p}_\col$, has the same magnitude in both equations but is directed in opposite direction according to the principle of {\it actio et reactio}.
This allows to derive the collision load $\vec{p}_\col$ by subtracting Eq. \eqref{eq:relVelFinal2} from Eq. \eqref{eq:relVelFinal1} resulting in 
\begin{equation}{\label{eq:relVelFinal3}}
\vec{p}_\col = \vec{K}_{12}^{-1} \cdot (\vec{u}^{n}_{\mathrm{r}} - \vec{u}^{n-1}_{\mathrm{r}} - \vec{u}_{\mathrm{ex,r}}) \hspace{0.1cm},
\end{equation}
with the total system matrix $\vec{K}_{12} = \vec{K}_1 + \vec{K}_2$ and $\vec{u}^{n}_{\mathrm{r}} = \vec{u}^{n}_{\mathrm{c,2}}-\vec{u}^{n}_{\mathrm{c,1}}$ the relative velocity of the particles at the contact point after the collision at time $t^n$. Since $\vec{u}^{n}_{\mathrm{r}}$ is still unknown, the Poisson hypothesis \citep{Lubarda2009} is used here to close this equation.  
It relates the normal relative velocity $\vec{u}^n_\mathrm{r,n}$ before and after the collision via
\begin{equation}\label{eq:edn}
\vec{u}^n_\mathrm{r,n} = -e_\mathrm{d,n}\, \vec{u}^{n-1}_\mathrm{r,n} = -e_\mathrm{d,n}\, (\vec{u}^{n-1}_\mathrm{r}\cdot\vec{n}) \, \vec{n} \hspace{0.1cm} ,
\end{equation}
where $e_\mathrm{d,n}$ is the dry coefficient of restitution. 
Analogously, the tangential velocities can also be related through a tangential restitution coefficient, $e_\mathrm{d,t}$, using
\begin{equation}\label{eq:edt}
\vec{u}^n_\mathrm{r,t} = -e_\mathrm{d,t}\, \vec{u}^{n-1}_\mathrm{r,t} = -e_\mathrm{d,t}\, (\vec{u}^{n-1}_\mathrm{r}\cdot\vec{t}) \, \vec{t} \hspace{0.1cm} 
\end{equation}
and the decomposition $\vec{u}^{n}_{\mathrm{r}} = \vec{u}^n_\mathrm{r,n}+\vec{u}^n_\mathrm{r,t}$. The unit tangent vector, $\vec{t}$, indicating the tangential projection of the particle motion is defined as $\vec{t}$ = ${(\vec{u}^{n-1}_\mathrm{r} - \vec{u}^{n-1}_\mathrm{r,n})}/{\lvert (\vec{u}^{n-1}_\mathrm{r} - \vec{u}^{n-1}_\mathrm{r,n})\rvert}$. 

Eqs. \eqref{eq:edn} and \eqref{eq:edt} allow to calculate the momentum transfer $\vec{p}_\col$ in Eq. \eqref{eq:relVelFinal3} from the initial state of the colliding particles at first touch, i.e. at time level $t^{n-1}$.
However, the tangential part of the relative velocity after the collision, $\vec{u}^{n}_\mathrm{r,t}$, remains an unknown quantity, since it depends upon whether the particles slide or stick to each other during the contact. In case of sticking, $\vec{u}^{n}_\mathrm{r,t}$ can be calculated using \eqref{eq:edt}, given that $e_\mathrm{d,t}$ is known {\em a priori}. In case of sliding, $\vec{u}^{n}_\mathrm{r,t}$ must be provided in a different way. 

In the present work, this is done by following the recommendations of \citet{Guendelman2003}.
As a first guess, it is assumed that the colliding particles stick fully at the contact point in tangential direction, i.e.~$e_\mathrm{d,t} = 0$. 
Using the Poisson hypothesis in normal direction with $e_\mathrm{d,n}$, the change in the relative velocity during a collision is
\begin{equation}	{\label{eq:Delta_u}}
	\Delta{\vec{u}} = \vec{u}^{n-1}_\mathrm{r} - \vec{u}^n_\mathrm{r} = \vec{u}^{n-1}_\mathrm{r} +\, e_\mathrm{d,n}\, (\vec{u}^{n-1}_\mathrm{r}\cdot\vec{n}) \, \vec{n} \hspace{0.1cm} ,
\end{equation}
so that Eq. \eqref{eq:relVelFinal3} turns into
\begin{equation}\label{eq:pc_stick}
    \vec{p}_\col = -\vec{K}_{12}^{-1} \cdot \Delta\vec{u} \; -\vec{K}_{12}^{-1} \cdot \vec{u}_{\mathrm{ex,r}} \hspace{0.1cm} .
\end{equation}
Here, the first term on the right hand side imposes the required velocity constraint of the Poisson hypothesis, commonly used in dry hard sphere collision models. The second term captures the momentum transferred by external loads during the collision process, and is usually not considered by common hard sphere models. 

In a second step of the collision modeling, $\vec{p}_\col$ is used to verify whether the particles stick or slide along each other during contact. 
According to the Coulomb friction model, the particle surfaces stick if $\lvert \vec{p}_\col \cdot \vec{t} \rvert \le \mu_s \, \lvert \vec{p}_\col \cdot \vec{n} \rvert$ with $\mu_s$ being the static coefficient of friction. Once this condition is fulfilled, $\vec{p}_\col$ according to Eq. \eqref{eq:pc_stick} is a good approximation of the collision response. Otherwise, $\vec{p}_\col$ has to be modified to allow sliding in tangential direction. Based on the Coulomb law of friction, $\vec{p}_\col$ during sliding can be expressed in the local coordinate system as
\begin{subequations}
	\begin{align}
	\vec{p}_\col &= p_\mathrm{n}\, \vec{n} + p_\mathrm{t} \,\vec{t}\\
		\label{eq:pc_sliding_2}
	&= p_\mathrm{n}\,(\vec{n} + \mu_\mathrm{k}\,\vec{t}) \, ,
	\end{align}
\end{subequations}
with the normal and tangential components $p_\mathrm{n}$ and $p_\mathrm{t}$, respectively. The factor $\mu_k$ is the kinetic coefficient of friction. It relates the tangential and normal part of the collision response via $p_\mathrm{t} = \mu_\mathrm{k} \, p_\mathrm{n}$.
The alteration of the tangential part of $\vec{p}_\col$ in Eq. \eqref{eq:pc_sliding_2} violates the Poisson hypothesis in normal direction. This is because $\vec{p}_\col$ in Eq. \eqref{eq:pc_stick} is derived by assuming that the particles stick to each other at the contact point, i.e. $\vec{u}^n_\mathrm{r,t} = 0$. Therefore, the normal momentum $p_\mathrm{n}$ in case of sliding has to be recalculated by inserting $\vec{p}_\col$ from Eq. \eqref{eq:pc_sliding_2} in Eq. \eqref{eq:pc_stick}
\begin{equation}
\left[\vec{K}_{12} \cdot p_\mathrm{n}\,(\vec{n} + \mu_\mathrm{k}\,\vec{t})\right] \cdot \vec{n} = \left[-\Delta\vec{u} -\vec{u}_{\mathrm{ex,r}}\right] \cdot \vec{n} \:  \hspace{0.1cm},
\end{equation}
so that
\begin{equation}
p_\mathrm{n} = - \; \frac{ \Delta\vec{u} \!\cdot\! \vec{n} \, + \, \vec{u}_{\mathrm{ex,r}} \!\cdot\! \vec{n} }{ \vec{K}_{12} \cdot \,(\vec{n} + \mu_\mathrm{k}\,\vec{t}) \cdot \vec{n} } \hspace{0.1cm} .
\end{equation}
Finally, the vectorial momentum transfer during a collision in case of tangential sliding is obtained from Eq. \eqref{eq:pc_sliding_2} and the associated torque via $\vec{l}_\col = \vec{r}_{\mathrm{c}} \times \vec{p}_\col $.

\subsection{Multiple simultaneous collisions}
\label{LCP}
When simulating sediment transport with thousands of particles, almost each single particle is in contact with several other particles at the same collision time. As a result, the collisional momentums, $\vec{p}_\col$ and $\vec{l}_\col$, must be an outcome of all the collisions. Using the approach in Section \ref{sec:single_collision}, the collision loads can only be calculated for a single collision between two particles. The derivation of multiple simultaneous collisions differs from the above and is considerably more extensive. To overcome this issue, a common assumption is to suppose that all particles collide in pairs, i.e. a single collision of two particles, one after another. This is also done here. Each single collision is modeled exactly as described above. Then, all single collisions are executed multiple times in an iterative procedure until $\vec{p}_\col$, acting between each single pair, prevents an interpenetration of the particles. This is the Signorini condition, i.e. $\vec{p}_\col \cdot \vec{n} \ge 0$. In each iteration step, the order of the collision pairs is varied at random to achieve the most realistic collision response of the sediment.
More details on the collision modeling of multiple simultaneous collisions can be found in \cite{Tschisgale2018b} including a pseudocode of the iterative procedure.
\section{{\label{sec:lubrication_model}}Lubrication model}
\subsection{Literature review}
When the distance between two approaching particles becomes small right before the particle surfaces actually touch, the fluid between them is squeezed out of the gap. This process takes place on small length and time scales but is important for the collision as a whole since it influences, e.g., the rebound height due to the related dissipation \citep{Davis1986}. In the IBM framework as used in this study the fluid motion in the liquid film between the particles is resolved as long as the gap between the particles is covered by at least two to three grid cells. When the gap between the particles becomes smaller, the fluid forces acting on the particles are no more resolved and are underpredicted \citep{kempe2012a, izard2014}. This is generally compensated by using an analytical expression for the lubrication force in addition to the fluid forces computed at the immersed boundary. This so-called lubrication model provides the correct inter-particle force, as done in the cited references. Here, the distance where this takes place is denoted $d^*_\mathrm{lub}$ with $d^*_\mathrm{lub} = 2\Delta_x$, as used by \citet{kempe2012a} .

Various lubrication models have been proposed in the literature to model the phases of the collision process with small surface distances before and after the phase of direct surface contact \citep{kempe2012a,  izard2014, biegert2017}. They are all based on the lubrication theory of \citet{brenner1961} and \citet{cox1967} and differ very little with respect to each other. For example, the model for spherical particles used by \citet{kempe2012a} reads
\begin{equation}\label{eq:lub_kempe}
\vec{f}^*_\mathrm{lub} = 
	\begin{cases} 
	0\hspace{0.1cm},   & \qquad \lvert\vec{d}\rvert > d^*_\mathrm{lub} \\
	\displaystyle -\frac{6\pi\mu_\mathrm{f}\vec{u}_{r,n}}{\lvert\vec{d}\rvert}\left( \frac{r_\mathrm{1}r_\mathrm{2}}{r_\mathrm{1}+r_\mathrm{2}}\right) ^2 \hspace{0.1cm}, & \qquad \eta \leqslant \lvert\vec{d} \rvert \leqslant d^*_\mathrm{lub} \\
	0 \hspace{0.1cm},   & \qquad \lvert\vec{d}\rvert < \eta \ \hspace{0.1cm},
	\end{cases}
\end{equation}
where $r_i$ is the radius of the $i$-th particle, $\vec{u}_{r,n}$ the normal component of the relative velocity, and $\mu_\mathrm{f} = \nu_\mathrm{f}\rho_\mathrm{f}$ the dynamic viscosity. The arguments of $\vec{f}^*_\mathrm{lub}$ are the independent variables for given fluid and particle pairing. For distances $\lvert \vec{d} \rvert > d^*_\mathrm{lub}$, the particle motion is resolved by the background Eulerian grid and no modeling of the lubrication forces is applied. Only if the surface distance is between $ \eta \leqslant \lvert\vec{d} \rvert \leqslant d^*_\mathrm{lub}$, the lubrication model sets in. The parameter $\eta$ constitutes a lower bound for the distance in Eq. \eqref{eq:lub_kempe}. It is motivated from a physical point of view by the roughness each and every surface in reality has. It is also needed from an algorithmic point of view since $\vec{f}^*_\mathrm{lub}$ diverges for $\lvert\vec{d} \rvert \to 0$.

The models used by \citet{izard2014}, and \citet{biegert2017} apply a constant force in the range $0 < \lvert\vec{d}\rvert < \eta$ instead of $\vec{f}^*_\mathrm{lub} = 0$ as in Eq. \eqref{eq:lub_kempe}. Moreover, the value of $\eta$ is different in each of these models. In the work of \citet{biegert2017} the parameter, $\eta$, is calibrated such that the rebound trajectory of a $\SI{6}{mm}$ steel particle after collision with a glass wall at Stokes number, $St_\mathrm{r} = 27$ matches with the experiments of \citet{gondret2002}. This results in a value of $\eta$ approximately $30$ times the experimental value $\eta = 10^{-4}D_\mathrm{eq}$ given in \citet{gondret2002} and used by \citet{kempe2012a}. \citet{jeffrey1982} proposed an extended version of the lubrication model which could be applied in the region $ \eta \leqslant \lvert\vec{d} \rvert \leqslant d^*_\mathrm{lub}$. This model includes extra terms to calculate the force. Nevertheless, this force is still inversely proportional to the separation distance.
\begin{figure*}[tp]
	\centering
	\setlength{\unitlength}{1cm}
	\begin{picture}(12,8)
	\put(0,0){\includegraphics{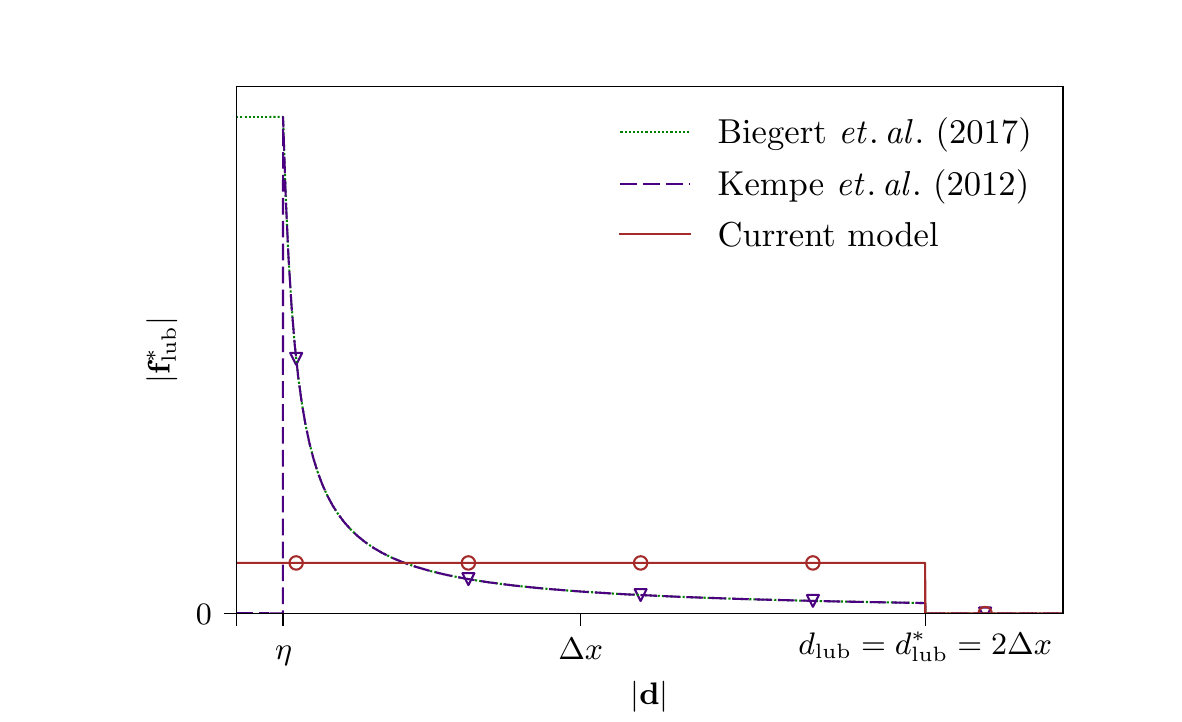}}
	\end{picture}
	\caption{Lubrication force applied in traditional models compared to the new model proposed here. The value employed for $\eta$, not drawn to scale, is different in the two functions, used by \citet{kempe2012a} and \citet{biegert2017}. In this graph, it is taken to be the same for simplicity. The circles and triangles identify distances assumed in successive time steps in case of a CFL number equal to $0.5$, as discussed in the text.}
	\label{fig:lub_models_sample}
\end{figure*}
The inverse proportionality to the distance $\lvert\vec{d}\rvert$, as in Eq. \eqref{eq:lub_kempe}, leads to a substantial variation in the force applied on the particle depending on where the particle happens to land at a particular time step. The issue is highlighted by the triangles in Fig. \ref{fig:lub_models_sample} corresponding to the evaluation of the lubrication force in case of a constant time step and a constant particle velocity, such that the CFL number is $0.5$. This choice is made for illustration here and can be different in applications where smaller CFL number is usually employed. Similar cases are obtained by shifting the entire set of evaluations by some value in $\lvert \vec{d} \rvert$-direction. The evaluation at discrete times, i.e. distances, samples the curves. It is obvious that if the particle approaches such that the first triangle in the diagram is shifted slightly towards the left, the last triangle will move to a much larger value so that the sum of the forces becomes different if Eq. \eqref{eq:lub_kempe} is used. Furthermore, it is also obvious that the modification for $\lvert\vec{d}\rvert < \eta$, which is not drawn to scale here, does not have much impact, as it makes the body thicker by a tiny amount only ($\eta \approx 10^{-4} D_\mathrm{eq}$) in the \citet{kempe2012a}. As a solution to this problem, \citet{biegert2017} suggested a sub-stepping scheme in the lines of \citet{Kidanemariam2014a} computing the particle motion with smaller timesteps than the fluid motion by dividing a fluid timestep into $150$ substeps, for example.
\begin{figure*}[tp]
	\centering
	\setlength{\unitlength}{1cm}
	\begin{picture}(16,7.5)
	\put(0,0){\includegraphics{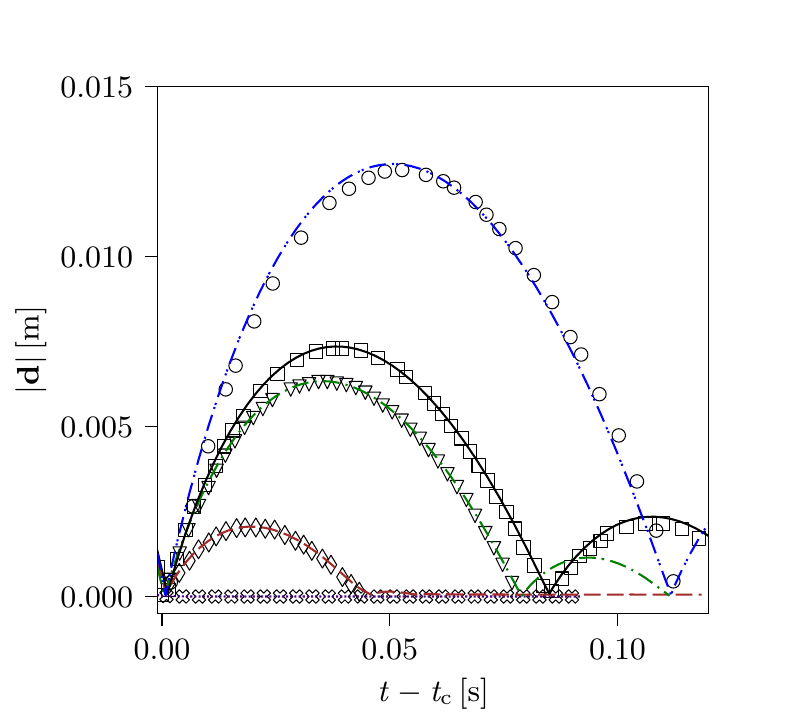}}
	\put(8,0){\includegraphics{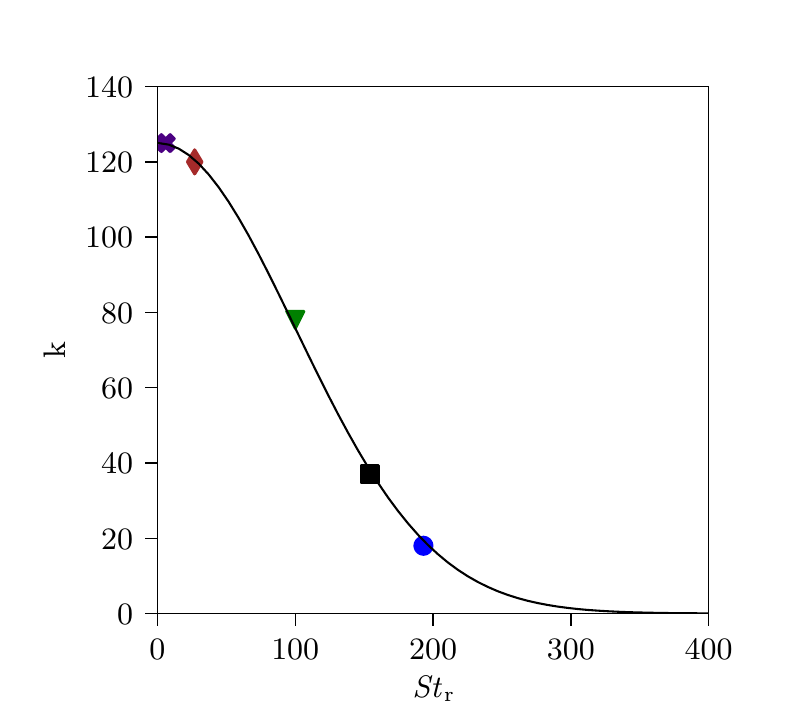}}
	\put( 0. , 7 )     {a) }
	\put( 8. , 7 )     {b) }  
	\end{picture}
	\caption{Calibration of the lubrication model to match the trajectories provided by \cite{gondret2002}. a) Particle trajectories after a normal collision with a wall in viscous fluid obtained in the experiments (symbols) and in the simulations (lines). Indigo: $St_\mathrm{r}=6$; brown: $St_\mathrm{r}=27$; green: $St_\mathrm{r}=100$; black: $St_\mathrm{r}=152$; blue: $St_\mathrm{r}=193$. Here, $t_\col$ is the time of collision. b) The lubrication model parameter $k$ as a function of $St_\mathrm{r}$. The color code of the symbols corresponds to the same values of $St_\mathrm{r}$ as in a). The continuous line is the fit according to Eq. \eqref{eq:k_st}.  }
	\label{fig:lub_calib}
\end{figure*}

However, in a sediment transport simulation, where tens of thousands of particles are colliding in one timestep, this method might prove inefficient and may increase the computational cost. \citet{Kidanemariam2014a} conducted sediment transport simulations with sub-stepping for the particle motion but used a simple collision model without any lubrication force. The coefficients of stiffness and damping in their soft-sphere collision model were calibrated to meet the experiments of \citet{gondret2002} and \citet{joseph2001}. Also, the dry coefficient of restitution was set to a value much smaller than the value for glass which is close to the one for quartz sand employed by \citet{gondret2002} and, thus, relevant for sediment transport. 

\subsection{New lubrication model}
To avoid the uncontrolled variation of $\vec{f}^*_\mathrm{lub}$ just before collision noticed by \citet{biegert2017} without sub-stepping, a different approach is proposed here. It consists of using a constant corrective force for all distances $ 0 < \lvert\vec{d} \rvert \leqslant d_\mathrm{lub} $, as shown in Fig. \ref{fig:lub_models_sample}, where $d_\mathrm{lub} $ is a predefined model parameter. It defines the distance range over which the lubrication force is applied and is much larger than $\eta$ in the models discussed above. The expression of this lubrication force is inspired \eqref{eq:lub_kempe}. This classical model contains factors reflecting geometry, approach velocity, viscosity, inverse distance, and a scalar constant. The first three are retained, $\lvert\vec{d} \rvert$ is replaced with $d_\mathrm{lub}$, and a new dependence on the Stokes number is introduced. The model is given by
\begin{equation}
	\vec{f}_\mathrm{lub} = 
	\begin{cases} 
		0\hspace{0.1cm},   & \qquad \lvert\vec{d}\rvert > d_\mathrm{lub} \\

		\displaystyle - k(St_\mathrm{r})\, \frac{\mu_\mathrm{f}\vec{u}_{r,n}}{d_\mathrm{lub}}\left( \frac{r_\mathrm{1}r_\mathrm{2}}{r_\mathrm{1}+r_\mathrm{2}}\right) ^2\hspace{0.1cm}, & \qquad \lvert\vec{d} \rvert \leqslant d_\mathrm{lub} \hspace{0.1cm}.
	\end{cases}
	\label{eq:lub_model}
\end{equation}
\cite{kempe2012a} argued that the fluid forces influencing the particle motion are well resolved as long as the distance between two particles, or between a particle and a wall is greater than $2\Delta x$. Following this recommendation, others have also applied the lubrication force in the region $ 0 \leqslant \lvert\vec{d} \rvert \leqslant 2\Delta_x$ \citep{izard2014,Ardekani2016,biegert2017}. The same value is used here setting $d_\mathrm{lub}=2\Delta x$. Furthermore, it will be reported in Sec. \ref{sec:sensitivity_analysis} below that the model is fairly insensitive to this parameter. 

The function $k(St_\mathrm{r})$ is determined such that the rebound trajectories match the experimental data of \citet{gondret2002}. The trajectories available in this reference, except those at very large $St_\mathrm{r}$, were used to calibrate the present model. This is accomplished using the relative Stokes number
\begin{equation}
	St_\mathrm{r} = \frac{1}{9}\frac{\rho_\mathrm{p}}{\rho_\mathrm{f}}\frac{\lvert \vec{u}_{r,n} \rvert}{\nu_\f} \left(\frac{r_\mathrm{1}r_\mathrm{2}}{(r_\mathrm{1}+r_\mathrm{2})/2}\right) \hspace{0.1cm}.
	\label{eq:rel_stokes}
\end{equation}
\textcolor{red}{Unlike the classical Stokes number for a single particle in a flow, which is based on the velocity difference between particle and fluid, this Stokes number is based on the difference between the velocities of the two particles, more precisely the component in the direction of the distance vector. This velocity scale was selected because it is the relevant scale for the lubrication forces. Also, the length scale $r_\mathrm{1}r_\mathrm{2}/((r_\mathrm{1}+r_\mathrm{2})/2)$, which is $r$ in the case of equal radii $r_\mathrm{1}=r_\mathrm{2}=r$, seems appropriate to address lubrication forces. In fact these scales already appear in the classical lubrication models looking like Eq. \eqref{eq:lub_kempe}. Finally, this choice is backed \textit{a posteriori} by the success of this parametrization achieved in covering the experimental data.} The dependence of the collision force on $St_\mathrm{r}$ is a new feature compared to previous models and introduces the particle density in the collision model.

The experiments of \citet{gondret2002} were conducted with spherical particles of different materials impacting on a glass wall in different fluids. In the present study a wall is represented as a sphere of radius $\SI{e+12}{m}$  and density $\SI{e+12}{kg/m^3}$. Replacing $r_\mathrm{1}$ with the radius of the colliding particle and $r_\mathrm{2}$ with $\SI{e+12}{m}$ in Eq. \eqref{eq:rel_stokes} yields the same definition of the Stokes number as used in the cited experiments. 
\subsection{Dependency of the new lubrication model on the Stokes number}

To provide a closed form expression of $k(St_\mathrm{r})$ to be used in Eq. \eqref{eq:lub_model}, a set of simulations of particle-wall collision were carried out. The physical parameters of particles and fluids as well as the dimensionless numbers characterizing the simulations are assembled in Tab. \ref{tab:lub_calibration}. The computational domain of size $6.6D_\mathrm{p}\times13.4D_\mathrm{p}\times6.6D_\mathrm{p}$ was discretized with $128\times256\times128$ cells in all the simulations. This gives a spatial resolution of $D_\mathrm{p}/\Delta_x=19$. Periodic boundary conditions were imposed in the wall-parallel directions and a no-slip and a free slip condition on the bottom and the top boundary, respectively. A fixed time step was used corresponding to $\mathrm{CFL}=0.1$ based on the terminal velocity of the particle. The initial condition was still fluid and a particle position at a vertical distance of $10D_\mathrm{p}$ above the wall.
\begin{table*} [htp]
	\centering
	\begin{tabular}{cccccllcl}
		\hline \hline
		Case	& $D_\mathrm{p}\,[\mathrm{m}]$	& $\rho_\mathrm{p}\,[\mathrm{kg\,m^{-3}}]$	& $\rho_\mathrm{f}\,[\mathrm{kg\,m^{-3}}]$ & $\nu_\mathrm{f}\,[\mathrm{m^2\,s^{-1}}]$	& $St_\mathrm{r}$ & $Re_\mathrm{p}$ & $e_\mathrm{d,n}$ 	& $k$\\
		\hline
		$1$ 	& $3\times10^{-3}$ 	& $7800$ 	& $965$ & $1.0363\times10^{-4}$ 	& $6$    & $6$   & $ 0.97 $ & $125$\\
		$2$ 	& $6\times10^{-3}$ 	& $7800$ 	& $965$ & $1.0363\times10^{-4}$ 	& $27$   & $30$  & $ 0.97 $ & $120$\\
		$3$ 	& $4\times10^{-3}$ 	& $7800$ 	& $953$ & $2.0986\times10^{-5}$ 	& $100$  & $110$ & $ 0.97 $ & $78$ \\
		$4$ 	& $3\times10^{-3}$ 	& $7800$ 	& $935$ & $1.0695\times10^{-5}$ 	& $152$  & $165$ & $ 0.97 $ & $37$ \\
		$5$ 	& $6\times10^{-3}$ 	& $7800$ 	& $953$ & $2.0986\times10^{-5}$ 	& $193$  & $212$ & $ 0.97 $ & $18$ \\
		\hline
		\hline
	\end{tabular}
	\caption{Material properties of particles and fluids used in the experiments of \citet{gondret2002} and in the present simulations. The values of the model parameter $k$ leading to matching of trajectories between simulation and experiment are listed as well. The particle Reynolds number, $Re_\mathrm{p}$, is defined using the particle velocity before collision and the particle diameter $D_\mathrm{p}$. }	
	\label{tab:lub_calibration}
\end{table*}

The experiments of \citet{gondret2002} were conducted such that the settling particle reached its terminal velocity before colliding. This would require a large enough computational domain or an artificial acceleration \citep{biegert2017}. In the simulations reported here, a moving frame methodology was used to accelerate the particle to its terminal velocity. Once the particle Stokes number of the experiment was attained, the moving frame was fixed and the particle was allowed to freely collide with the wall. 

The lubrication force was applied as discussed above. Each simulation was repeated varying the value of $k$ value until the rebound trajectory matched the experiment. The resulting trajectories are shown in Fig. \ref{fig:lub_calib}a. The value of $k$ resulting from this procedure is provided in Tab. \ref{tab:lub_calibration} and plotted in Fig. \ref{fig:lub_calib}b. It is found that $k$ decreases substantially with $St_\mathrm{r}$. \citet{izard2014} demonstrated that the effect of the lubrication force is negligible for $St_\mathrm{r} \geqslant 200$, supporting the present observation and indicating that the decreasing trend continues for values of $St_\mathrm{r}$ larger than the ones selected here. It turns out that the behavior of $k(St_\mathrm{r})$ can be fitted extremely well by a simple Gauss function which is also shown in Fig. \ref{fig:lub_calib}b. This exponential function is motivated by the correlation between Stokes number and the effective coefficient of restitution $e_\mathrm{n}$ conceptualized by \citet{izard2014}. These authors proposed a model to reproduce the dependency of the effective restitution coefficient on the Stokes number observed in experiments and numerical studies. The function used here is  
\begin{equation} {\label{eq:k_st}}
k(St_\mathrm{r}) = \alpha \exp\left( -\frac{1}{2\sigma^2} St_\mathrm{r}^2\right) \hspace{0.1cm}, 
\end{equation}
with $ \alpha = 125$, and $ \sigma = 100 $ obtained from the fit to the simulation data. 
\subsection{Extension to non-spherical particles}
\textcolor{red}{To model the lubrication force acting on non-spherical particles, particular geometrical properties of the contacting surfaces need to be considered in the model,  such as the two principal radii of curvature at the contact point \citep{Cox1974}. \cite{Janoschek2013}, proposed an accurate method to compute the lubrication corrections in a case of two approaching spheroids using a third-order description of the particular particle surfaces which requires the determination of the associated coefficients at each contact point. In a simulation with thousands of colliding ellipsoids, however, such an algorithm is computationally too expensive. As an alternative, \citet{Ardekani2016} proposed an expression of the lubrication force based on the Gaussian curvature of the spheroidal particles at the contact point, where the local particle surface is approximated by a sphere of the same curvature. Since this approach provides a good balance between accuracy and computational efficiency it is employed in the present work as well, just for the more general case of ellipsoids instead of spheroids.} The Gaussian curvature of an ellipsoid at a point $ (X,Y,Z) $ in a body-fixed frame is 
\begin{equation}
	\mathrm{G} = \displaystyle \frac{1}{a^2b^2c^2\left[ \frac{X^2}{a^4} + \frac{Y^2}{b^4} +\frac{Z^2}{c^4} \right]^2 } \hspace{0.1cm},
\end{equation}
and the radius of Gaussian curvature is given by \citep{gray1997}
\begin{equation}
R_\mathrm{G} = \frac{1}{\sqrt{\mathrm{G}}} \hspace{0.1cm}.
\label{eq:Rg}
\end{equation}
Along these lines the lubrication force between two non-spherical particles is determined here using Eq. \eqref{eq:lub_model} with the radius of the spherical particles, $r_\mathrm{1}$ and $r_\mathrm{2}$ replaced by $R_\mathrm{G,1}$ and $R_\mathrm{G,2}$ according to Eq. \eqref{eq:Rg}, respectively, which depend on the particular collision point. The Stokes number in Eq. \eqref{eq:rel_stokes} is also determined in this way, and the expression for $k$ in Eq. \eqref{eq:k_st} is employed with the same values of $\alpha$ and $\sigma$ as used for spherical particles.
\section{{\label{sec:sensitivity_analysis}}Sensitivity analysis}
\begin{table*} [tp]
	\centering
	\begin{tabular}{ccclclc}
		\hline \hline
		$D_\mathrm{eq}\,[\mathrm{m}]$	& $\rho_\mathrm{p}\,[\mathrm{kg\,m^{-3}}]$	& $\rho_\mathrm{f}\,[\mathrm{kg\,m^{-3}}]$ & $\nu_\mathrm{f}\,[\mathrm{m^2\,s^{-1}}]$	&  $e_\mathrm{d,n}$  &$\mu_\mathrm{s}$ & $\mu_\mathrm{k}$  \\
		\hline
		$3\times10^{-3}$ 	& $7800$ 	& $935$ & $1.0695\times10^{-5}$  & $0.97$ & $0.02$ &  $0.02$	\\
		\hline
		\hline
	\end{tabular}
	\caption{Physical parameters of the fluid and particle used in the sensitivity analysis.}	
	\label{tab:sensitivity_analysis1}
\end{table*}
\begin{table*} [tp]		
		\centering
		\begin{tabular}{l|cccc}
			\hline
			\hline
			Variable & \multicolumn{4}{c}{Value}  \\
			\hline
			$\mathrm{CFL}$	& $0.1$ 	& $ 0.3 $ &	 $\mathbf{0.6}$	& $0.9$ \\
			\hline
			$D_\mathrm{p}/\Delta_x$ & $10$ 	& $ \mathbf{20} $ &	$ 40 $	&	\\
			\hline
			$d_\mathrm{lub}$ & $1\Delta_x$ 	& $ \mathbf{2\Delta_x} $ &	$ 3\Delta_x $	& $4\Delta_x$ \\
			\hline
			$y_\mathrm{p,0}\, \mathrm{[m]}$ & $\mathbf{0.03}$ 	& $ 0.02995 $ &	$ 0.02990 $	& $0.02985$ \\
			\hline
			\hline
		\end{tabular}
	\caption{The four parameters that were varied in the sensitivity analysis. Reference values are written in boldface.}
	\label{tab:sensitivity_analysis2}
\end{table*}
The sensitivity of the proposed model to the $\mathrm{CFL}$ number, the spatial resolution, $D_\mathrm{p}/\Delta_x$, the width of the lubrication region, $d_\mathrm{lub}$, and the initial position of the falling particle, $y_\mathrm{p,0}$, was thoroughly tested. The last test was performed to address the possibility of a corresponding sensitivity pointed out by \citet{biegert2017}. The same computational domain as mentioned previously in Sec. \ref{sec:lubrication_model} was used for this purpose. A particle of diameter $\SI{0.003}{m}$ was positioned at $y_\p = \SI{0.0015}{m}$ close to the bottom  wall of the computational box. Another particle of the same diameter and heavy enough to settle on a straight vertical path was unleashed above the bottom particle at a position $y_\mathrm{p,0}$ to freely accelerate and to collide directly with the top of the fixed particle. The material properties and dimensionless numbers characterizing the flow are such that upon collision the Stokes number is $St_\mathrm{r} \approx 113$. Further data are assembled in Tab. \ref{tab:sensitivity_analysis1}. For $St_\mathrm{r} \geq 200$, the lubrication force has minimal effect on the rebound trajectory, whereas the collision is viscously damped and is inelastic for $St_\mathrm{r} < 10$ \citep{izard2014}. The chosen Stokes number is a compromise between both the situation assuring a rebound which is influenced by the lubrication model.

A total of $15$ simulations were conducted by varying only one parameter out of the four studied parameters at a time. The parameters being varied and their values are tabulated in Tab. \ref{tab:sensitivity_analysis2}. The set of reference values is marked with boldface. An additional simulation with much higher spatial resolution, $D_\mathrm{p}/\Delta_x=40$, and $\mathrm{CFL}=0.1$ was used as a reference. The maximum rebound height $H_\mathrm{m}$ was used to assess the trajectory of all simulations and to compare them to the reference case.  

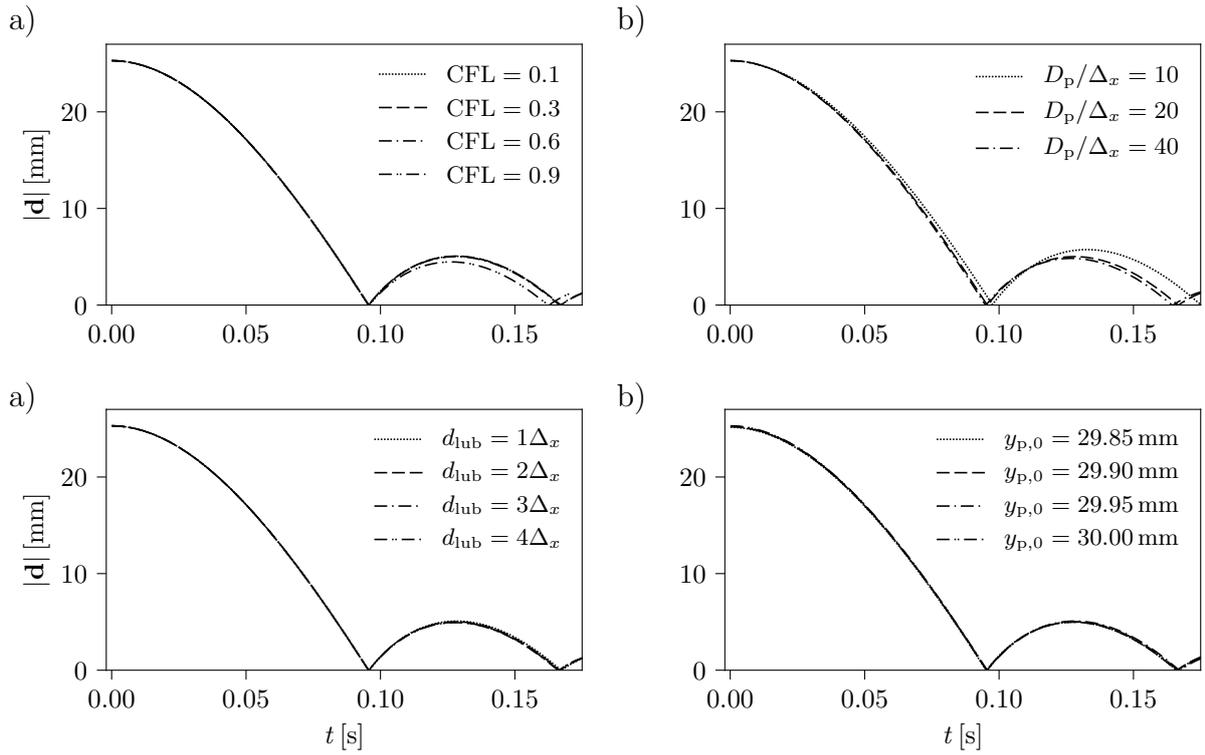
\begin{figure*}[tp]
	\centering
	\setlength{\unitlength}{1cm}
	\begin{picture}(16,10.5)
	\put(0,0){\includegraphics{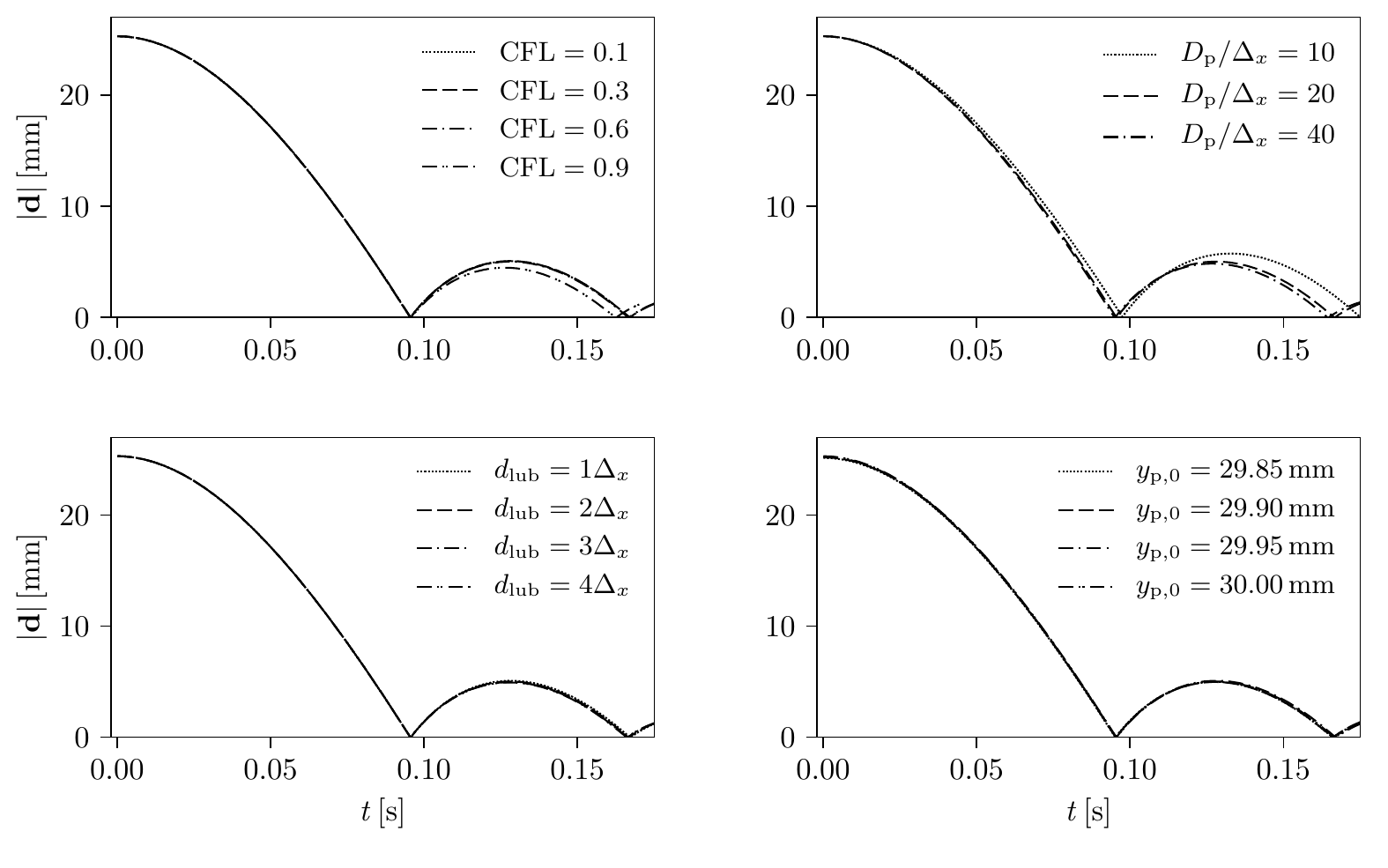}}
	\put( 0. , 10. )     {a) }
	\put( 8.0 , 10. )     {b) }  
	\put( 0. , 5. )     {a) }
	\put( 8.0 , 5. )     {b) }  
	\end{picture}
	\caption{Trajectories obtained in the sensitivity study defined by the parameters in Tab. \ref{tab:sensitivity_analysis2}. a) Variation of $\mathrm{CFL}$ number, b) variation of $D_\mathrm{p}/\Delta_x$, c) variation of $d_\mathrm{lub}$, and d) variation of $y_\mathrm{p,0}$. In each comparison the other three parameters are kept constant. }
	\label{fig:sensitivity}
\end{figure*}

The trajectories of various simulations are compared in Fig. \ref{fig:sensitivity}. For larger time steps, e.g. $\mathrm{CFL} = 0.9$, the trajectory turns out to be damped yielding a difference in $H_\mathrm{m}$ of $4\%$. In contrast, there is no visible difference in the trajectories for $\mathrm{CFL}\leqslant 0.6$ (Fig. \ref{fig:sensitivity}a). Reducing the spatial resolution from $D_\mathrm{p}/\Delta_x=20$ to $D_\mathrm{p}/\Delta_x=10$ gives a relative variation of $H_\mathrm{m}$ by $10\%$. In fact, fluid forces on a particle that is resolved with only $10$ cells are only marginally captured by the IBM. Doubling the resolution, the difference in $H_\mathrm{m}$ with respect to the reference case using the same resolution and $\mathrm{CFL} = 0.1$ decreases to only $2\%$, Fig. \ref{fig:sensitivity}b.

There is no noticeable difference in the trajectories when changing the lubrication zone $d_\mathrm{lub}$ (Fig. \ref{fig:sensitivity}c). In fact \eqref{eq:lub_model} was constructed so that an increase in $d_\lub$ increases the distance over which $\vec{f}_\lub$ is applied but also decreases $\vec{f}_\lub$ proportionally to $1/d_\lub$, so that the integral under the curve in Fig. \ref{fig:lub_models_sample} remains constant. This provides substantial robustness, as witnessed by this test. Finally, it was verified that no difference in the trajectories is obtained if the initial position of the falling particle is changed by a small amount (Fig. \ref{fig:sensitivity}d). This is expected since the lubrication force is constant over the entire region $0<\lvert\vec{d}\rvert<d_\lub$, so that the sampling effect highlighted in Fig. \ref{fig:lub_models_sample} for the classical model does not occur any more.

\section{Validation}
\subsection{Normal collision of a spherical particle with a wall}

The benchmark rebound trajectories provided by \citet{gondret2002} are used to calibrate the model. Therefore, they can be reproduced fairly well, as shown in Fig. \ref{fig:lub_calib}a. Many further cases were investigated in the cited experiment to determine the dependency of the restitution coefficient for a normal collision, 
\begin{equation}
e_\mathrm{n} =  -\frac{\lvert \vec{u}_\mathrm{p,n,out} \rvert}{\lvert \vec{u}_\mathrm{p,n,in} \rvert} \hspace{0.1cm},
\end{equation}
as a function of the Stokes number. Here, $\vec{u}_\mathrm{p,n,in}$ and $\vec{u}_\mathrm{p,n,out}$ are the input and output particle normal velocities measured just before and after the lubrication zone, i.e. when the surfaces have a distance of $d_\mathrm{lub}$ identified with start and end of the collision process. Both velocities are normal to the colliding wall. The influence of the viscous fluid surrounding the particle can be highlighted very well when using the restitution coefficient of dry collision $e_\mathrm{d,n}$ as a reference, which is done in Fig. \ref{fig:validation_sphere}a for the data of \citet{gondret2002} and \citet{joseph2001}. The experimental behavior of $e_\mathrm{n}/e_\mathrm{d,n}$ versus $St_\mathrm{r}$ is reproduced very well using this model. All previously developed collision models have been compared with this curve in the literature. \\

The same situation was simulated with the new collision model using the same fluid and particle properties as mentioned in Tab. \ref{tab:sensitivity_analysis1} and a computational domain of size $13D_\mathrm{p}\times26D_\mathrm{p}\times13D_\mathrm{p}$ discretized with $256\times512\times256$ cells. The particle was dropped from a distance $y_\mathrm{p,0} = \SI{0.06}{m}$ to the wall. The Stokes number was varied by successively changing the gravitational acceleration from $g=0.02\times\SI{9.81}{m/s^2}$ to  $g=20\times\SI{9.81}{m/s^2}$, similar to \citet{biegert2017}. The parameter values, $d_\mathrm{lub}=2\Delta_x$, $D_\mathrm{p}/\Delta_x \approx 20$, $\mathit{CFL} = 0.5$ were used in these tests. Fig. \ref{fig:validation_sphere}a shows that the dependency of the restitution coefficient on $St_\mathrm{r}$ obtained in the simulations with the proposed model is captured very well. Observing the considerable experimental scatter a better agreement can hardly by expected. 
\subsection{Oblique collision of a spherical particle with a wall}
\label{sec:sph_obl_col}
\begin{figure*}[tp]
	\centering
	\setlength{\unitlength}{1cm}
	\begin{picture}(16,8)
	\put(0,0){\includegraphics{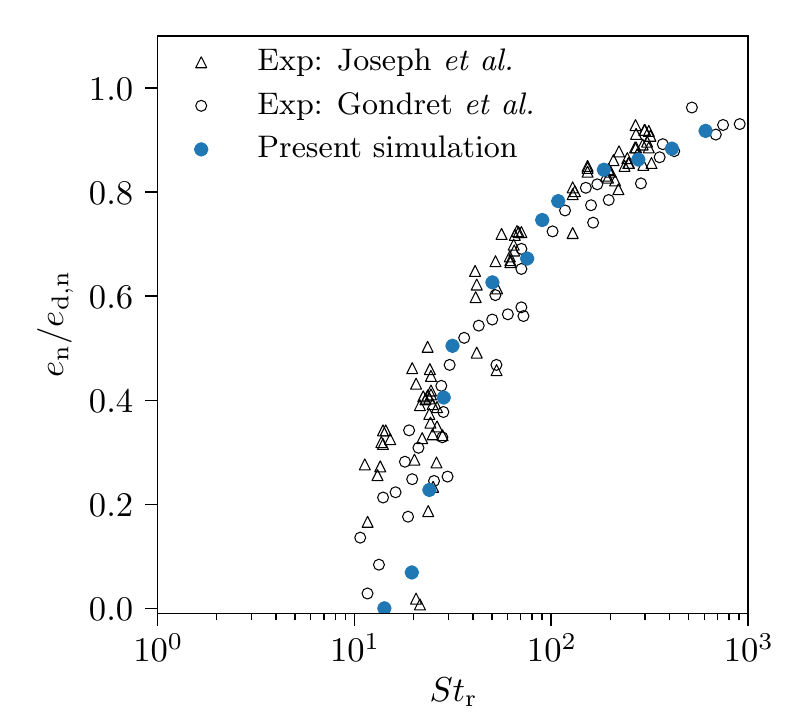}}
	\put(8,0){\includegraphics{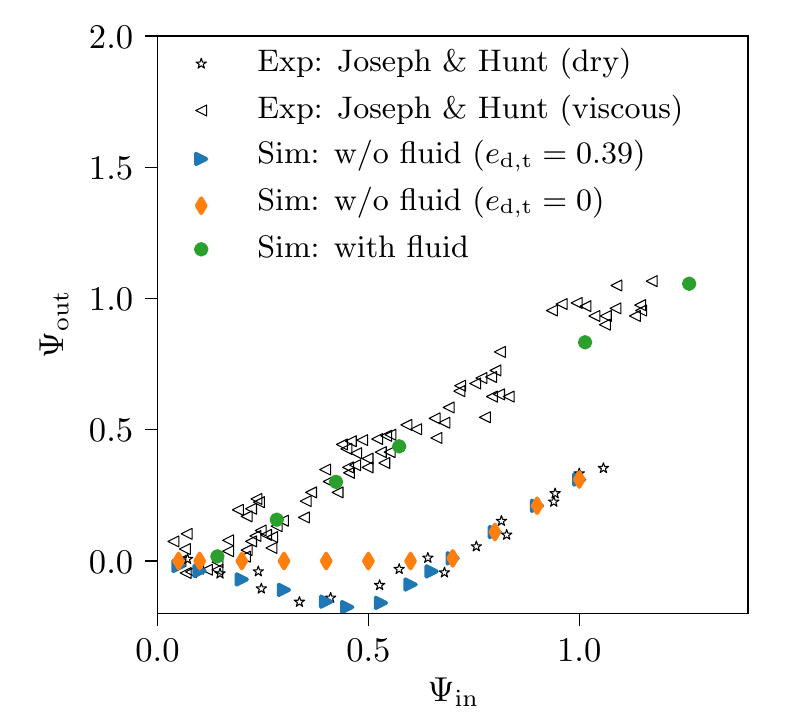}}
	\put( 0. , 7.5 )     {a) }
	\put( 8. , 7.5 )     {b) }  
	\end{picture}
	\caption{Results of the simulations conducted to validate the collision model. a) Normalized restitution coefficient plotted against the particle Stokes number. The present simulation results are compared with the experiments of \citet{gondret2002} and \citet{joseph2001}. b) Present simulations without and with fluid to validate the collision model for an oblique wall collision. The experimental data of \citet{joseph2004} are used for reference.}
	\label{fig:validation_sphere}
\end{figure*}
The model also covers tangential forces so that it is also validated for oblique particle-wall collisions in a dry system and in the presence of a viscous liquid. A computational domain of $26D_\mathrm{p}\times13D_\mathrm{p}\times13D_\mathrm{p}$ with resolution $D_\mathrm{p}/\Delta_x \approx 20$ was used in these simulations. The particle was initially placed at $(X,Y,Z)=(4D_\mathrm{p},7D_\mathrm{p},7D_\mathrm{p})$ with its motion is driven by the acceleration 
\begin{equation}
\vec{g} = (\,g \sin(\phi_\mathrm{g}),-g\cos(\phi_\mathrm{g}),0 \,)^T \hspace{0.1cm}.
\end{equation}

Here, $\phi_\mathrm{g}$ is the angle between the wall-normal axis of the computational grid and gravity. Since the distance between the starting point of the trajectory is relatively short and the trajectory prior to collision is a straight line, so that the incidence angle of the particle with the wall-normal $\phi_\mathrm{in}=\phi_\mathrm{g}$ in these simulations. The magnitude of the imposed acceleration equals $g=20\times\SI{9.81}{m/s^2}$ to achieve the high Stokes numbers of the experiments. \\

The simulations are compared with the experiments of \citet{joseph2004} on oblique particle-wall collisions in air and aqueous solutions of glycerol.
\begin{table*} [tp]
	\centering
	\begin{tabular}{ccclclcc}
		\hline \hline
		$D_\mathrm{eq}\,[\mathrm{mm}]$ & $\rho_\mathrm{p}\,[\mathrm{kg\,m^{-3}}]$	& $\rho_\mathrm{f}\,[\mathrm{kg\,m^{-3}}]$ & $\nu_\mathrm{f}\,[\mathrm{m^2\,s^{-1}}]$	&  $e_\mathrm{d,n}$&$\mu_\mathrm{s}$ & $\mu_\mathrm{k}$ & $e_\mathrm{d,t}$\\
		\hline
		$2.5$ & $2540$ 	& - & -  & $0.97$	& $0.15$    & $0.15$ & $0.0$   \\
		$2.5$ & $2540$ 	& - & -  & $0.97$	& $0.15$    & $0.15$  & $0.39$ \\
		$2.5$ & $7800$ 	& $998$ & $1.007\times10^{-6}$  & $0.97$	& $0.02$    & $0.02$  & $0.34$ \\
		\hline
		\hline
	\end{tabular}
	\caption{Particle and fluid properties used in the simulations to validate the new model for oblique collisions. The parameters are similar to the ones used in the experiments of \citet{joseph2004}. Cases without fluid properties provided refer to dry collisions.}	
	\label{tab:oblique_collision}
\end{table*}
\citet{joseph2004} compared the rebound angle $\Psi_\mathrm{out}$ to the impact angle $\Psi_\mathrm{in}$, where
\begin{equation}
\Psi_\mathrm{in} = \frac{\lvert \vec{u}_\mathrm{c,t,in} \rvert}{\lvert \vec{u}_\mathrm{c,n,in} \rvert} \hspace{0.1cm},  \, \, \,
\Psi_\mathrm{out} = \frac{\lvert \vec{u}_\mathrm{c,t,out} \rvert}{\lvert \vec{u}_\mathrm{c,n,in} \rvert} \hspace{0.1cm},
\end{equation}
so that, e.g., $\Psi_\mathrm{in} = \tan(\phi_\mathrm{in})$. Following \cite{joseph2004}, the velocities at the contact point are employed here, instead of the particle center, as previously used in the definition of $e_\mathrm{n}$ above. Since in these tests the spherical particle is not rotating significantly before the collision the velocity of the contact point, $\vec{u}_\mathrm{c,in}$ is the same as the velocities at the particle center,  $\vec{u}_\mathrm{p,in}$. However, the tangential velocity of the contact point after the collision differs from the center point velocity and is calculated as $\vec{u}_\mathrm{c,t,out} = (\vec{u}_\mathrm{c}\!\cdot\vec{t})\,\vec{t}$. \\

Fig. \ref{fig:validation_sphere}b displays the results of the simulations and provides a comparison with the experiments for the same configuration. The dry collision was first simulated under the assumption of full stick at the contact point i.e. $e_\mathrm{d,t} = 0$. The results are represented with the diamond symbols. They match very well when the particle slides. The negative values in the rolling mode are not captured properly, however, as expected. This is obtained by introducing $e_\mathrm{d,t} = 0.39$ in Eq. \eqref{eq:edt}, as proposed by \citet{joseph2004}. The match between the results with the new model and the reference data is extremely good so that the validation is successfully accomplished.

\subsection{Oblique collisions of ellipsoidal particles} \label{ssec:Coll_NS}
\begin{figure}[t]
	\centering
	\def\svgwidth{0.45\linewidth}
	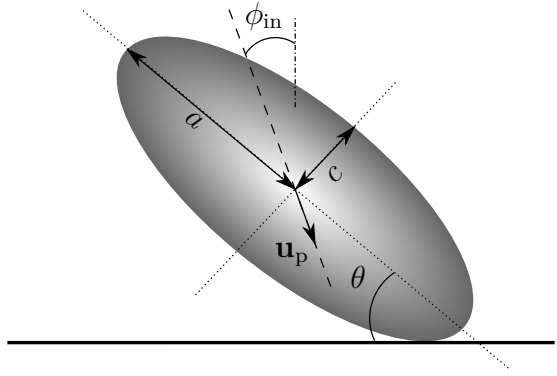
	\caption{Off-center collision of an ellipsoidal particle with a wall and definition of the particle orientation angle $\theta$.}
	\label{fig:par_orientation}	
\end{figure}

For non-rotating spherical particles colliding with a wall in viscous fluid the rebound only depends on the Stokes number i.e. $e_\mathrm{n}/e_\mathrm{d,n}$ = $f(St_\mathrm{r})$ \citep{gondret2002}. For ellipsoidal particles, even without rotation, other parameters such as the ratio of the particle axes and the particle orientation become pertinent in addition, i.e. $e_\mathrm{n}/e_\mathrm{d,n}$ = $\phi(St_\mathrm{r}, \, a/b, \, b/c, \, \theta/2\pi)$, as illustrated by Fig. \ref{fig:par_orientation}, so that the parameter space is substantially enlarged. If rotation is considered \textcolor{red}{in addition, as is done here}, $e_\mathrm{n}/e_\mathrm{d,n}$ is also a function of the particle angular velocity in case of ellipsoids because the magnitude of the normal velocity of the contact point is a sum of $\vec{u}_\p \cdot \vec{n}$ and $(\vec{\omega}_\p \times \vec{r}_{\mathrm{c}})\cdot \vec{n}$. The latter term vanishes for spherical particles. The orientation of an ellipsoidal particle just before collision, $\theta/2\pi$, and its angular velocity are difficult to control in an experiment. This is one of the reasons why, to the best of the authors knowledge, no experimental data similar to the ones presented for spherical particles by \cite{gondret2002} and \cite{joseph2001} are available with ellipsoids. \\

\begin{figure*}[tp]
	\centering
	\setlength{\unitlength}{1cm}
	\begin{picture}(16,8)
	\put(4,0){\includegraphics{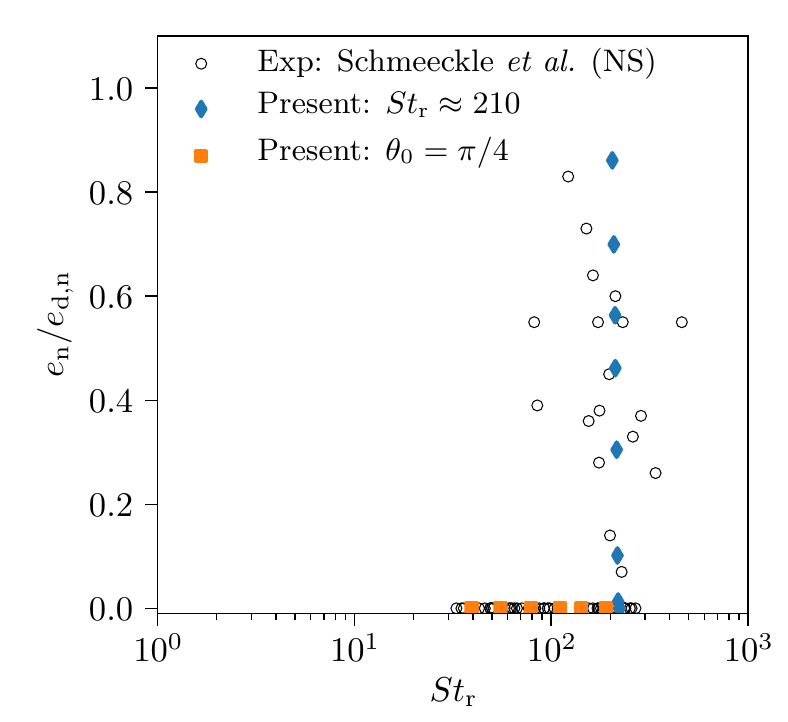}}
	\end{picture}
	\caption{Ratio of the restitution coefficient of the normal collision,  $e_\mathrm{n}$, to the coefficient for dry collision,  $e_\mathrm{d,n}$, obtained for collisions of an ellipsoidal particle with an inclined wall using the present model. Diamond: Simulations with the present model and the Stokes number fixed at $St_\mathrm{r} \approx 210$, using different initial particle orientations, $\theta_0$. Square: Fixed initial particle orientation, $\theta_0 = \pi/4$, variable Stokes number, as indicated by the horrizontal axis. The experimental data of \citet{schmeeckle2001} for collisions of natural sediment particles with an inclined glass wall are plotted for reference. }
	\label{fig:validation_NS}
\end{figure*}

As a remedy collisions of natural sediment particles are considered. These have been studied intensively in the literature on sediment transport, as this issue constitutes a fundamental aspect of bed-load transport. In one such study, \citet{schmeeckle2001} explored the collision of natural sediment particles with an inclined glass wall in water. They reported a total of $67$ such collisions and observed that a collision can be elastic or inelastic depending upon the orientation of the colliding particle. Only in $18$ collisions of these a rebound of the particle center was noticed. The restitution coefficient, $e_\mathrm{n}$, of the rebounding particles varied between $0.07$ and $0.83$ depending on their orientation just before the collision. The particle orientation $\theta$ ranged between $17\pi/45$ and $\pi/2$ with an average of $\theta = 9\pi/20$. The Stokes number was also different in all these collisions with the majority of them, $12$ out of $18$, happening at $St_\mathrm{r} \approx 210$, as seen in Fig. \ref{fig:validation_NS}. The $49$ particles which did not rebound, hence exhibiting $e_\mathrm{n} = 0$, collided with the wall at a Stokes number between $St_\mathrm{r}=33$ and $St_\mathrm{r}=268$. For these non-rebounding collisions, the average particle orientation was $\theta = 7\pi/20$ in the experiment. The data of these experiments for natural sediment are plotted in Fig. \ref{fig:validation_NS} for later comparison. \\

The physical properties of the particle and the fluid used in the simulations reported here were selected to match the physical properties of natural sediment i.e. quartz, and water (Tab. \ref{tab:NS_collision}). Choosing the particle shape for the simulations requires some selection to narrow down the huge parameter space. The major and minor half-axes of the natural sediment investigated were measured from the photographs by \textcolor{red}{\citet{schmeeckle2001}}, so that the information of only two axes is available, with the ratio of the major axis to the minor axis $b/c$ being essentially in the range of $1.4$ to $2.5$. On this background, the natural sediment particles were represented by oblate ellipsoids, i.e. $a=b$, and an aspect ratio of $b/c = 2$. \textcolor{red}{Beyond the restricted availability of data the choice of an oblate shape was motivated by the measurements of the shape of quartz sand particles conducted by \citet{Smith_Cheung2002}. They found a much higher percentage of oblate-like particles ($42\%$) compared to prolate-like ones ($14\%$), blades ($9\%$), and sphere-like particles ($35\%$), so that an oblate shape seems to be a good choice.} \\

\begin{table*} [tp]
	\centering
	\begin{tabular}{cccclclcc}
		\hline \hline
		$D_\mathrm{eq}\,[\mathrm{mm}]$ & $b/c$ & $\rho_\mathrm{p}\,[\mathrm{kg\,m^{-3}}]$	& $\rho_\mathrm{f}\,[\mathrm{kg\,m^{-3}}]$ & $\nu_\mathrm{f}\,[\mathrm{m^2\,s^{-1}}]$	&  $e_\mathrm{d,n}$&$\mu_\mathrm{s}$ & $\mu_\mathrm{k}$ & $e_\mathrm{d,t}$\\
		\hline
		$6$ & $2$ & $2650$ 	& $998$ & $1.007\times10^{-6}$  & $0.97$	& $0.15$    & $0.15$  & $0.39$ \\
		\hline
		\hline
	\end{tabular}
	\caption{Particle and fluid properties employed for the simulations of an oblate particle colliding with an inclined glass wall.}	
	\label{tab:NS_collision}
\end{table*}

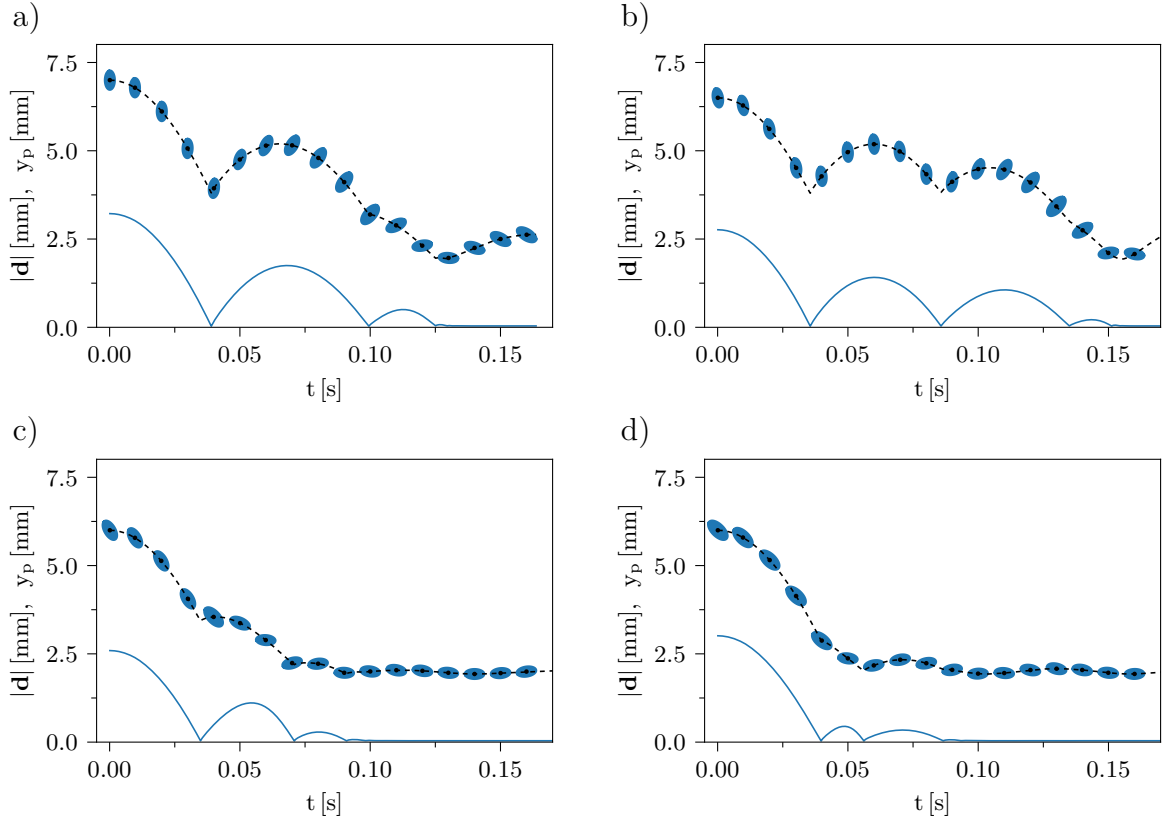
\begin{figure*}[tp]
	\centering
	\setlength{\unitlength}{1cm}
	\begin{picture}(16,11)
	\put(0,5.5){\includegraphics{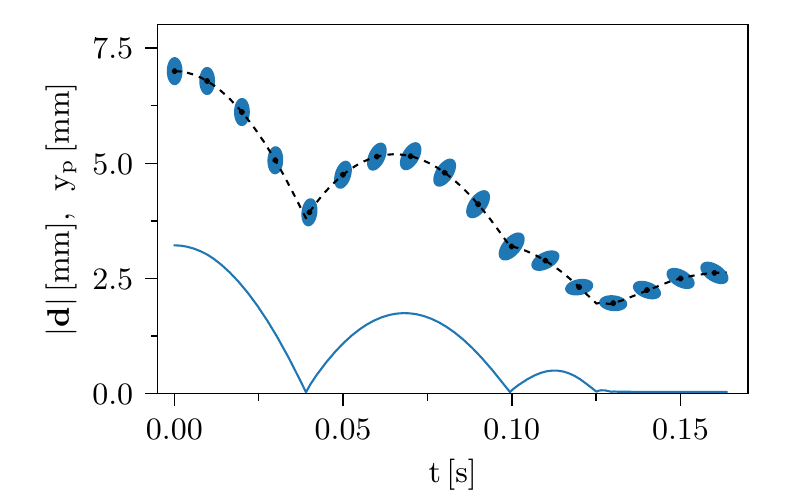}}
	\put(8,5.5){\includegraphics{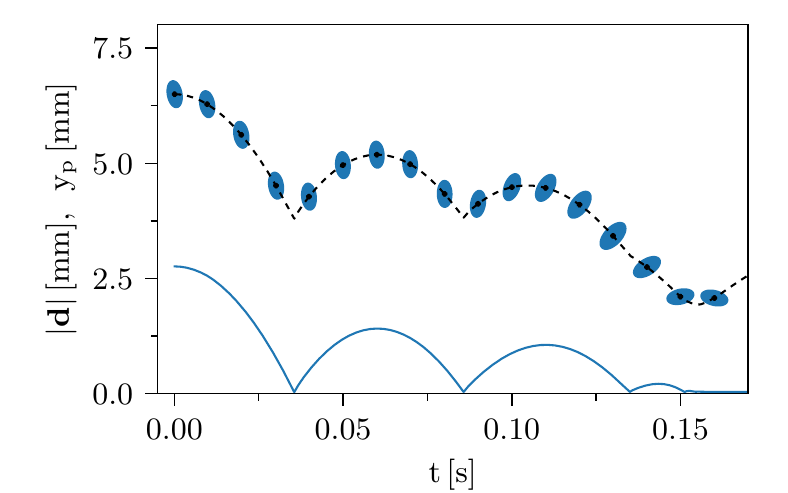}}
	\put(0,0){\includegraphics{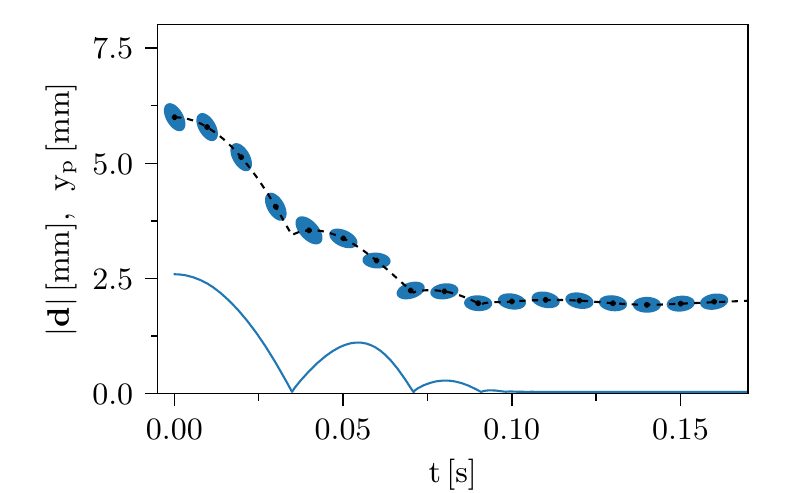}}
	\put(8,0){\includegraphics{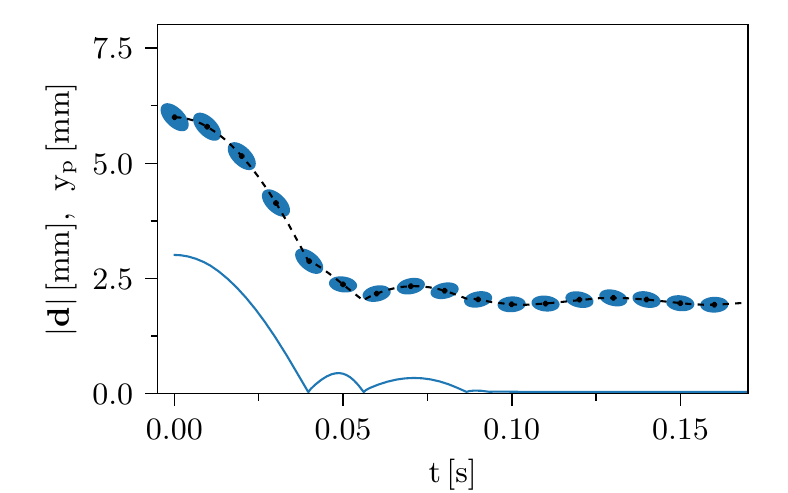}}
	\put( 0.5 , 10.5 )     {a) }
	\put( 8.5 , 10.5 )     {b) }  
	\put( 0.5 , 5. )     {c) }  
	\put( 8.5 , 5. )     {d) }  	
	\end{picture}
	\caption{Oblique collisions of an oblate ellipsoid ($b/c=2$) with a wall at $\phi_\mathrm{in} = \pi/6$. The Stokes number in all cases is $St_\mathrm{r} \approx 210$. Only the initial orientation on the particle was changed. a) $\theta = \pi/2$, b) $\theta = 4\pi/9$, c) $\theta = \pi/3$, and d) $\theta = \pi/4$. The dotted line represents the wall normal position of the particle center and the solid line shows the distance between the particle surface and the wall. Ellipses, not to scale, drawn with time interval of $\SI{0.01}{s}$ visualize the orientation of the particle. }
	\label{fig:oblate_traj}
\end{figure*}

A computational domain of size $\SI{0.04}{m} \times \SI{0.04}{m} \times \SI{0.04}{m}$ discretized with $256 \times 256 \times 256$ cells was used here. Since the majority of the collisions observed in the experiment occurred out at an incidence angle of $\phi_\mathrm{in} \approx \pi/6$, the same value was used in the simulations as well. In a first set of runs, the particle was dropped from a height adjusted in each case so as to reach the desired Stokes number, $St_\mathrm{r} \approx 210$. This height depends on the initial orientation and is about $y_\mathrm{p,0} \approx 1.15\,D_\mathrm{eq}$ (Fig. \ref{fig:oblate_traj}). A few other simulations (not shown here) were conducted with higher $y_\mathrm{p,0}$ but with the same Stokes number, orientation, and angular velocity before collision to understand the effect of the particle wake on the effective restitution coefficient. It was found that the ratio $e_\mathrm{n}/e_\mathrm{d,n}$ changed only very little. This is due to the short time interval between the states $\vec{u}_\mathrm{p,in}$ and $\vec{u}_\mathrm{p,out}$ not providing enough time for the wake to influence this process, even more for the Stokes numbers considered here. The rebound trajectories after the collision were different, though, but this is beyond the collision process to be modeled here and an entire topic on its own.

The simulations were conducted with different initial orientations $\theta_0$ of the particle and $\vec{\omega}_\mathrm{p,0} = 0$. It turned out that during the approach the particle orientation changed only very little, less than five degrees, so that $\theta \approx \theta_0$. The resulting restitution coefficients are reported in Fig. \ref{fig:validation_NS}. If the center of the oblate particle after the collision continues to approach the wall instead of rebounding, as the collision point, the restitution coefficient is negative. In such a case, $e_\mathrm{n}$ was set to zero here, as in the experiment. It was observed in the simulations that the particle does not rebound when $\theta \approx \pi/4$. To investigate this further, a second set of simulations was conducted with $\theta_0 = \pi/4$ varying the gravitational acceleration $g$ to achieve different Stokes numbers. The results are included in Fig. \ref{fig:validation_NS} as well. As a result of the simulations conducted here, it can be seen that the collision model reproduces the phenomenon observed in \cite{schmeeckle2001} very well. \\

The initial and boundary conditions of these simulations are well specified. To provide a reference for later investigations and an illustration of the computed solution four sample trajectories are shown in Fig. \ref{fig:oblate_traj} obtained for a Stokes number $St_\mathrm{r} \approx 210$ and different initial orientations $\theta_0 = [\pi/2, 4\pi/9, \pi/3, \pi/4]$. In every sub-figure, there are two curves. The variation of the distance between the surfaces over time is plotted with a solid line and the wall normal position of the particle center is shown using a dashed line. Additionally, the particle orientation is visualized every $\SI{0.1}{s}$ using ellipses. In Fig. \ref{fig:oblate_traj}a and \ref{fig:oblate_traj}b, a clear rebound of both the particle center as well as the contact point can be noticed. In case of initial orientation $\theta_0=\pi/3$, shown in Fig. \ref{fig:oblate_traj}c, though, the particle center rebounds, but the restitution coefficient is much smaller and equals $0.125$. If $\theta_0=\pi/4$, the center of the particle continues to approach the wall although there is a rebound of the contact point. The center of the particle, then rebounds after the second collision. 

\subsection{Normal collisions of oblate particles with a wall}
To explore the effect of the particle orientation and the ratio of the particle axes further, more simulations were conducted. Here, a setup similar to the experiments of \citet{gondret2002} is used with the parameters listed in Tab. \ref{tab:sensitivity_analysis1}. The same computational domain as in Sec. \ref{ssec:Coll_NS} was employed with $\phi_\mathrm{in} = 0$. The equivalent diameter of the ellipsoid equals $D_\mathrm{eq} = \SI{6}{mm}$ in the simulations. The change in the particle orientation before collision is negligible due to the high density ratio, $\rho_\p/\rho_\f = 7.8$, taken over from the experiments with spheres and the short distance, so that $\theta = \theta_0$. The particle shape was varied by changing the ratio $b/c$ while keeping $a/b=1$ and $D_\mathrm{eq}$ unchanged. Three different ratios were used, $b/c=1$, $b/c=1.5$, and $b/c=2.0$, with increasing flatness $(a+b)/2c$ according to the definition of \cite{Cailleux1945}. 

\begin{figure*}[tp]
	\centering
	\setlength{\unitlength}{1cm}
	\begin{picture}(16,7.5)
	\put(0,0){\includegraphics{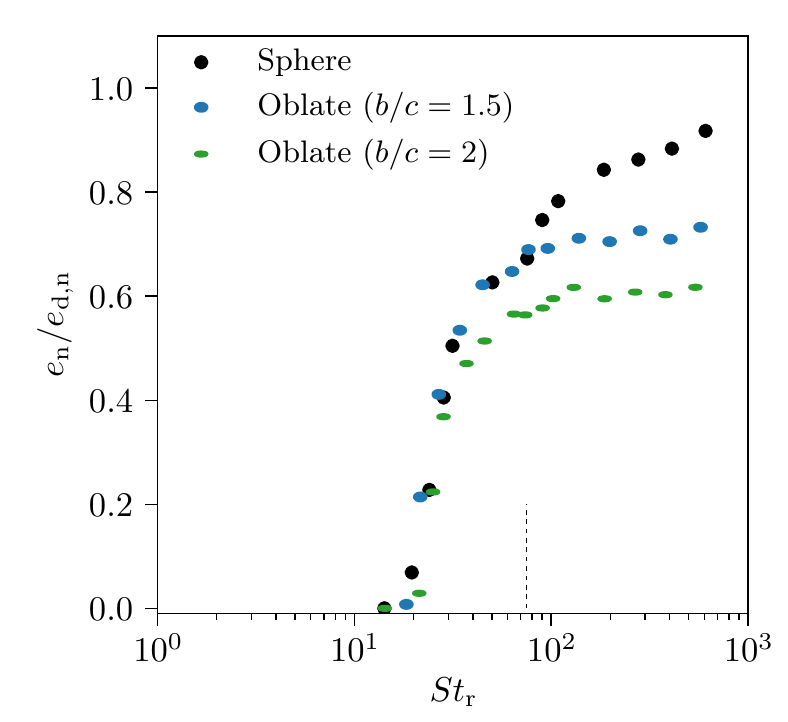}}
	\put(8,0){\includegraphics{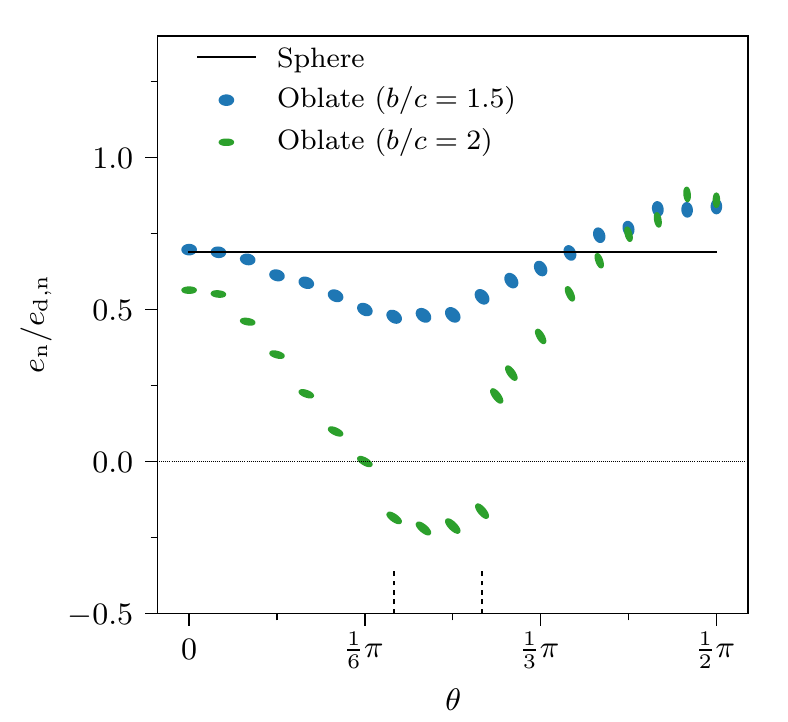}}
	\put( 0. , 7.5 )     {a) }
	\put( 8. , 7.5 )     {b) }  
	\end{picture}
	\caption{Normal collision of particles of different shape with a wall showing the dependency of the relative restitution coefficient on the Stokes number, the particle orientation, and the particle shape. a) The Stokes number and the particle shape are varied for a constant orientation. b) The particle orientation and its shape are changed keeping the Stokes number unchanged, $St_\mathrm{r} \approx 75$, indicated by a dashed line in a). The vertical dashed lines in sub-figure b) are positioned at the values $\theta \approx 7\pi/36$ and $\theta \approx 10\pi/36$ mentioned in the text. The symbols give a hint on the particle orientation upon collision and the particle flatness.}
	\label{fig:oblate_flatWall}
\end{figure*}
Two sets of simulations were undertaken for the three shapes considered. One was conducted with $\theta = 0$ varying the Stokes number by changing $\lvert \vec{g}\rvert$. Fig. \ref{fig:oblate_flatWall}a shows the results. It is observed that for oblate particles with $\theta = 0$, in contrast to spherical particles, the relative restitution coefficient, $e_\mathrm{n}/e_\mathrm{d,n}$, increases with increasing Stokes number and saturates at some value below $1$. This maximum value decreases with increasing flatness of the particle, $e_\mathrm{n}/e_\mathrm{d,n}\approx 0.70$ for $b/c=1.5$ and $e_\mathrm{n}/e_\mathrm{d,n}\approx 0.60$ for $b/c=2.0$. This is in line with the expectation, since lubrication forces become larger with increasing radius of curvature. Furthermore, the added mass effect, generating a force opposite to accelerations, changes with particle shape and increases with aspect ratio when an oblate particle is accelerated along its shorter axis. 

Fig. \ref{fig:oblate_flatWall}b shows the change in the restitution coefficient when changing the particle orientation prior to the contact. This is investigated for the three particle shapes examined. Here, the Stokes number was kept constant at a value of about $St_\mathrm{r} \approx 75$. It can be seen that the particle orientation upon contact, $\theta$, has a huge influence on the restitution coefficient. For $b/c=1.5$, the value of $e_\mathrm{n}/e_\mathrm{d,n}$ decreases with increasing $\theta$ up to $\theta \approx 7\pi/36$. It stays roughly constant between $\theta \approx 7\pi/36$ and $\theta \approx \pi/4$ before increasing to its maximum value at $\theta \approx 17\pi/36$. Interestingly, for $\theta$ around $\pi/2$, the restitution coefficient attains values higher than for a spherical particle. This is due to reduced lubrication because of the smaller radius of curvature at the contact point and due to the reduced added mass effects for this orientation of the particle with respect to its trajectory.

The observed trends are similar in the case $b/c=2.0$ and are more pronounced. The restitution coefficient has negative values between $\theta \approx \pi/6$ and $\theta \approx 10\pi/36$. A negative value of $e_\mathrm{n}$ means that the velocity of the particle centre after the collision is negative and the particle center still continues to approach the wall, even if there is a rebound of the contact point. This leads to further collisions with the wall. After two or three rebounds, the particle comes into an enduring contact with the wall still rolling back and forth before eventually coming to rest. 

\textcolor{red}{Let us finally mention that a small number of simulations with prolate particles were conducted in a similar way as reported in Fig. \ref{fig:oblate_flatWall} showing the same qualitative behaviour.}

\section{Conclusions}
A numerical method to model the collision between two arbitrary ellipsoids in a viscous media was developed which is conceptualized based on the hard-sphere collision model. First, an efficient and accurate contact detection algorithm was presented that exploits the fact that the surface tangent vectors at the contact points are perpendicular to the distance vector. The iterative procedure to detect the shortest surface distance between two ellipsoids converges fast and does not include any parameter that needs to be adjusted. 

With the commonly used lubrication models to capture the fluid forces acting on the colliding particles when the gap is too small to be resolved in the DNS-IBM framework these forces are inversely proportional to the distance. This can yield inconsistency in the rebound trajectories of the colliding particles since the lubrication force, then, is sensitive to the time step chosen. To avoid such an uncontrolled variation in the force and to improve the efficiency of the model, a new lubrication model was presented. In this model a constant force is applied in the region $0 < \lvert \vec{d} \rvert \leq d_\mathrm{lub}$. Making this force a function of the Stokes number is a new feature introduced here. 

A main achievement of the new method proposed here is the inclusion of fluid, lubrication, and gravitational forces directly in the IBM when determining the momentum transfer during an impact, which in other methods is generally handled as a dry collision, instead. The complete model was validated thoroughly against experimental data for normal and oblique collisions of a spherical particle with a wall, and against experimental sediment data for the oblique collision of an ellipsoidal particle with a wall. 

Using well specified initial conditions, the effects of particle orientation and shape on the normal coefficient of restitution $e_\mathrm{n}$ were investigated. Spherical particles tend to achieve the maximum value of $e_\mathrm{n}$, i.e. the restitution coefficient in dry environment $e_\mathrm{d,n}$, at higher Stokes numbers. In contrast, the ratio $e_\mathrm{n}/e_\mathrm{d,n}$ decreases significantly as the flatness of the particle increases. Particles of the same shape colliding at a particular Stokes number but with different orientation have different rebounds. The particle center may even not rebound at all, although the contact point rebounds. This effect is more pronounced with flatter particles. In the case of an oblate particle with axes ratio $b/c = 2.0$, for example, the normal restitution coefficient has even negative values for some orientations. 

Altogether, the new approach to model the collision between ellipsoidal particles in viscous media provides excellent agreement with the experiments. The new contact detection algorithm and the lubrication model are efficient and easy to implement. Furthermore, the setup used to investigate the binary collision of an oblate spheroid with the wall is well reproducible in a laboratory experiment as well as in simulations. The results obtained can be used for validation purposes. In the future, the collision model shall be used to study the effect of the particle shape on bed-load transport in case of small and large particle loading.

\newcommand{\sectionbreak}{\clearpage}

\section*{Acknowledgements}
	All the simulations presented here were carried out at the Center for Information Services and High Performance Computing (ZIH), TU Dresden, Germany. The authors thank ZIH for providing the computational resources and support. RJ gratefully acknowledges the scholarship provided by the Land Sachsen, No. L-201535. TS was employed by the Institute of Fluid Mechanics, TU Dresden at the time this study was carried out.

\bibliographystyle{abbrvnat}
\bibliography{../Dissertation/literatur}{}

\end{document}